\documentclass[prc,twocolumn,showpacs,twoside,showpacs]{revtex4}

\usepackage{amssymb}
\usepackage{graphicx}
\usepackage{dcolumn}
\usepackage{bm}
\usepackage{float}
\emergencystretch=\maxdimen
\begin{document}

\title{Spectroscopy of excited states of unbound nuclei $^{30}$Ar and $^{29}$Cl}
\author{X.-D. Xu$^{1,2,3}$}
\author{I. Mukha$^{3}$}
\author{L.V. Grigorenko$^{4,5,6}$}
\author{C. Scheidenberger$^{2,3}$}\thanks{Corresponding author: christoph.scheidenberger@physik.uni-giessen.de}
\author{L. Acosta$^{7,8}$}
\author{E. Casarejos$^{9}$}
\author{V.~Chudoba$^{4}$}
\author{A.A. Ciemny$^{10}$}
\author{W. Dominik$^{10}$}
\author{J. Du\'{e}nas-D\'{\i}az$^{11}$}
\author{V. Dunin$^{12}$}
\author{J. M. Espino$^{13}$}
\author{A. Estrad\'{e}$^{14}$}
\author{F.~Farinon$^{3}$}
\author{A. Fomichev$^{4}$}
\author{H. Geissel$^{2,3}$}
\author{T.A. Golubkova$^{15}$}
\author{A. Gorshkov$^{4,16}$}
\author{Z. Janas$^{10}$}
\author{G. Kami\'{n}ski$^{17,13}$}
\author{O.~Kiselev$^{3}$}
\author{R. Kn\"{o}bel$^{3,2}$}
\author{S. Krupko$^{4,16}$}
\author{M. Kuich$^{18,10}$}
\author{Yu.A. Litvinov$^{3}$}
\author{G. Marquinez-Dur\'{a}n$^{11}$}
\author{I. Martel$^{11}$}
\author{C. Mazzocchi$^{10}$}
\author{C. Nociforo$^{3}$}
\author{A.K. Ord\'{u}z$^{11}$}
\author{M. Pf\"{u}tzner$^{10,3}$}
\author{S. Pietri$^{3}$}
\author{M. Pomorski$^{10}$}
\author{A. Prochazka$^{3}$}
\author{S. Rymzhanova$^{4}$}
\author{A.M. S\'{a}nchez-Ben\'{\i}tez$^{11}$}
\author{P. Sharov$^{4}$}
\author{H. Simon$^{3}$}
\author{B. Sitar$^{19}$}
\author{R. Slepnev$^{4}$}
\author{M. Stanoiu$^{20}$}
\author{P. Strmen$^{19}$}
\author{I. Szarka$^{19}$}
\author{M. Takechi$^{3}$}
\author{Y.K. Tanaka$^{3,21}$}
\author{H. Weick$^{3}$}
\author{M. Winkler$^{3}$}
\author{J.S. Winfield$^{3}$}
\affiliation{$^1$School of Physics and Nuclear Energy Engineering, Beihang University, Beijing 100191, China}
\affiliation{$^2$II. Physikalisches Institut, Justus-Liebig-Universit\"{a}t Gie{\ss}en, Gie{\ss}en 35392, Germany}
\affiliation{$^3$GSI Helmholtzzentrum f\"{u}r Schwerionenforschung GmbH, Darmstadt 64291, Germany}
\affiliation{$^4$Flerov Laboratory of Nuclear Reactions, JINR, Dubna 141980, Russia}
\affiliation{$^5$National Research Nuclear University ``MEPhI'', Moscow 115409, Russia}
\affiliation{$^6$National Research Centre ``Kurchatov Institute'', Kurchatov sq. 1, Moscow 123182, Russia}
\affiliation{$^7$INFN, Laboratori Nazionali del Sud, Via S. Sof\'{\i}a, Catania 95123, Italy}
\affiliation{$^8$Instituto de F\'{\i}sica, Universidad Nacional Aut\'{o}noma de M\'{e}xico, M\'{e}xico, D.F. 01000, Mexico}
\affiliation{$^9$University of Vigo, Vigo 36310, Spain}
\affiliation{$^{10}$Faculty of Physics, University of Warsaw, Warszawa 02-093, Poland}
\affiliation{$^{11}$Department of Applied Physics, University of Huelva, Huelva 21071, Spain}
\affiliation{$^{12}$Veksler and Baldin Laboratory of High Energy Physics, JINR, Dubna 141980, Russia}
\affiliation{$^{13}$Department of Atomic, Molecular and Nuclear Physics, University of Seville, Seville 41012, Spain}
\affiliation{$^{14}$University of Edinburgh, EH1 1HT Edinburgh, United Kingdom}
\affiliation{$^{15}$Advanced Educational and Scientific Center, Moscow State University, Moscow 121357, Russia}
\affiliation{$^{16}$FSBI ``SSC RF ITEP'' of NCR ``Kurchatov Institute'', Moscow 117218, Russia}
\affiliation{$^{17}$Institute of Nuclear Physics PAN, Krak\'{o}w 31-342, Poland}
\affiliation{$^{18}$Faculty of Physics, Warsaw University of Technology, Warszawa 00-662, Poland}
\affiliation{$^{19}$Faculty of Mathematics and Physics, Comenius University, Bratislava 84248, Slovakia}
\affiliation{$^{20}$IFIN-HH, Post Office Box MG-6, Bucharest, Romania}
\affiliation{$^{21}$University of Tokyo, Tokyo 113-0033, Japan}
%

\begin{abstract}
Several states of proton-unbound isotopes $^{30}$Ar and $^{29}$Cl were investigated by measuring their in-flight decay products, $^{28}$S+proton+proton and $^{28}$S+proton, respectively. A refined analysis of $^{28}$S-proton angular correlations indicates that the ground state of $^{30}$Ar is located at $2.45^{+0.05}_{-0.10}$ MeV above the two-proton emission threshold. The theoretical investigation of the $^{30}$Ar ground state decay demonstrates that its mechanism has the transition dynamics with a surprisingly strong sensitivity of the correlation patterns of the decay products to the two-proton decay energy of the $^{30}$Ar ground state and the one-proton decay energy as well as the one-proton decay width of the $^{29}$Cl ground state. The comparison of the experimental $^{28}$S-proton angular correlations with those resulting from Monte Carlo simulations of the detector response illustrates that other observed $^{30}$Ar excited states decay by sequential emission of protons via intermediate resonances in $^{29}$Cl. Based on the findings, the decay schemes of the observed states in $^{30}$Ar and $^{29}$Cl were constructed. For calibration purposes and for checking the performance of the experimental setup, decays of the previously-known states of a two-proton emitter $^{19}$Mg were remeasured. Evidences for one new excited state in $^{19}$Mg and two unknown states in $^{18}$Na were found.
\end{abstract}
\maketitle

\section{Introduction}\label{sec:introduction}

Two-proton (2\emph{p}) radioactivity is an exotic nuclear decay mode resulting in the simultaneous emission of two protons. It was proposed for the first time by Goldansky in the early 1960s~\cite{Goldansky1960NP}. In this pioneering work, simultaneous two-proton emission was predicted to appear in the even-proton number ($Z$) isotopes beyond the proton drip-line, in which one-proton (1\emph{p}) emission is energetically prohibited but the ejection of two protons is energetically allowed due to the pairing interaction. More than 40 years after its prediction, ground-state $2p$ radioactivity was discovered in 2002~\cite{Pfutzner2002EPJA,Giovinazzo2002PRL}. Two experiments independently observed that the ground state (g.s.) of $^{45}$Fe decays by simultaneous emission of two protons. Later $^{54}$Zn~\cite{Blank2005PRL}, $^{19}$Mg~\cite{Mukha2007PRL}, $^{48}$Ni~\cite{Pomorski2011PRC}, and $^{67}$Kr~\cite{Goigoux2016PRL} were found to be other g.s.\ 2\emph{p} radioactive nuclei.

Among the g.s.~2\emph{p} emitters hitherto observed, the half-lives of $^{45}$Fe, $^{48}$Ni, and $^{54}$Zn are in the range of several ms, which can be accessed by the conventional implantation-decay method. In the case of $^{19}$Mg, whose half-life was predicted in the range from a few ps to a few ns~\cite{Grigorenko2003PRC}, a technique based on particle tracking of decays in flight (see details in Ref.~\cite{Mukha2010PRC}) was applied in order to investigate its decay properties. In this experiment, the trajectories of 2\emph{p} decay products of $^{19}$Mg were measured by double-sided silicon micro-strip detectors. The 2\emph{p}-decay vertices and fragment correlations were reconstructed. The 2\emph{p} decay energy and half-life of $^{19}$Mg g.s.~were determined, which represented the first case of 2\emph{p} radioactivity in $s$-$d$~shell nuclei~\cite{Mukha2007PRL}. In a recent work~\cite{Voss2014PRC}, the half-life of 2\emph{p} decay of $^{19}$Mg g.s.~was measured by another experimental technique, the extracted half-life value is consistent with the first measurements.

After the discovery of 2\emph{p} radioactivity several theoretical efforts were dedicated to predictions of the 2\emph{p} radioactivity landscape. In a systematic study of lifetime dependencies on the decay energy and three-body correlations applied to a number of isotopes by using a three-body model~\cite{Grigorenko2003PRC}, dozens of prospective true 2\emph{p} emitters were predicted. Among these candidates, $^{19}$Mg, $^{48}$Ni, and $^{54}$Zn have been proven to be indeed true 2\emph{p} emitters. In a recent study, the global landscape of g.s.~2\emph{p} radioactivity has been quantified by the energy density functional theory~\cite{Olsen2013PRL}. The main conclusion of this work is that 2\emph{p}-decaying isotopes exist in almost every isotopic chain between elements Ar and Pb, which indicates that g.s.~2\emph{p} radioactivity is a typical feature for the proton-unbound isotopes with even atomic numbers. Those theoretical predictions provide guidance for the experimental search of 2\emph{p} radioactive nucleus. For instance, $^{30}$Ar was predicted to be an $s$-$d$ shell true 2\emph{p} emitter by the three-body model~\cite{Grigorenko2003PRC}. The prediction for the $^{30}$Ar g.s.\ half-life was $T_{1/2}(^{30}\text{Ar}) = 0.7~\text{-}~33$ ps and the predicted separation energies were $S_{2p}(^{30}\rm{Ar})=-1.43$ MeV and $S_p(^{30}\rm{Ar})=0.35$ MeV, respectively. Considering its short lifetime, the in-flight decay method was applied. Several states in $^{30}$Ar and its 1\emph{p}-decay daughter nucleus $^{29}$Cl were investigated. The observation of $^{30}$Ar and $^{29}$Cl low-lying states, including their g.s., was reported in Ref.~\cite{Mukha2015PRL}. The assigned ground and first excited states of $^{29}$Cl were found at $1.8^{+0.1}_{-0.1}$ MeV and $2.3^{+0.1}_{-0.1}$ MeV above the $1p$ threshold, respectively. The g.s.~of $^{30}$Ar was found to be at $2.25_{-0.10}^{+0.15}$ MeV above the 2\emph{p} emission threshold~\cite{Mukha2015PRL}. A sophisticated data analysis together with theoretical investigations revealed that the g.s.~of $^{30}$Ar is located at $2.45^{+0.05}_{-0.10}$ MeV above the 2\emph{p} emission threshold~\cite{Golubkova2016PLB}. Due to a strong Thomas-Ehrman shift, the lowest states in $^{30}$Ar and $^{29}$Cl point to a violation of isobaric mirror symmetry in the structure of these unbound nuclei. Detailed investigations of the decay mechanism of the $^{30}$Ar ground state show that it is located in a transition region between simultaneous 2\emph{p} decay and sequential emission of protons. Such an interplay between the true three-body and the sequential two-body decay mechanism is the first-time observation for nuclear ground state decays. For the first excited $2^+$ state of $^{30}$Ar, the hint on so-called fine structure in the 2\emph{p} decay was obtained by detecting two decay branches either into the ground state or first excited state of $^{28}$S~\cite{Mukha2015PRL}.

Besides the g.s.\ and first excited states, several higher-lying excited states of $^{30}$Ar and $^{29}$Cl were also populated in this experiment. The present manuscript describes the details of the experiment and reports the first spectroscopy of observed excited states. The structure of the present manuscript is organized in the following way. To begin with, the experimental setup is introduced with the emphasis on the employed special ion-optic settings. Then the nuclear structure information on several observed states of the known 2\emph{p} emitter $^{19}$Mg is presented. Afterwards, the detailed analysis of the angular correlations between decay products and the extraction of the decay properties of several excited states of $^{30}$Ar as well as $^{29}$Cl is described. Finally, discussions on the transition dynamics of $^{30}$Ar g.s.\ decays and the decay mechanisms of observed excited states of $^{30}$Ar are presented.

\section{Experiment}\label{sec:Experiment}

The $^{30}$Ar experiment was performed at the Fragment Separator (FRS)~\cite{Geissel1992NIMB} at GSI (Darmstadt, Germany).
The FRS was operated with ion-optical settings in a separator-spectrometer mode. The primary 885 MeV/u $^{36}$Ar beam with an intensity up to $2\times10^{9}$ $s^{-1}$ impinged on a 8 $\rm{g/cm}^2$ $^9$Be production target. The 620 MeV/u $^{31}$Ar fragments with an average intensity of 50 $s^{-1}$ were selected as a secondary beam and transported by the first half of the FRS to bombard a $^9$Be reaction target located at the middle focal plane F2 of the FRS. The thickness and the transverse dimension of the reaction target is 4.8 $\rm{g/cm}^2$ and $5\times 5~\rm{cm}^2$, respectively. At the first focal plane F1 of the FRS, an aluminum wedge degrader was installed in order to achieve an achromatic focusing of $^{31}$Ar at the reaction target. $^{30}$Ar nuclei were produced via one-neutron (1n) knockout from the $^{31}$Ar ions. The decay products of $^{30}$Ar were tracked by a double-sided silicon micro-strip detector array placed just downstream of the reaction target. The projectile-like outgoing particles from the reaction target were analyzed by the second half of the FRS, which was operated as a magnetic spectrometer. The magnet settings between the focal planes F2 and F4 were tuned to transmit the targeted heavy ion (HI) fragments (e.g., $^{28}$S) down to the last focal plane F4.

\begin{figure}[!htbp]
\centerline{\includegraphics[scale=0.21, angle=0]{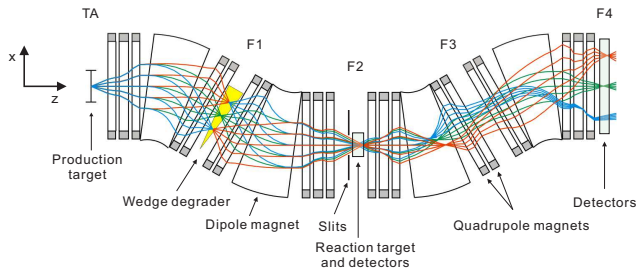}}
\caption{(Color online) Scheme of the FRS ion-optical system. The colored lines represent the calculated trajectories of $^{31}$Ar ions. The box at F2 denotes the experimental station including reaction target and tracking detectors. The horizontal slits at F2 are displayed. Detectors for the particle identification are represented by the box at F4. See text for details.}
\label{ion-optic_scheme}
\end{figure}

The above-mentioned operation mode of the FRS requires a special ion-optical setting. Since the transverse dimensions of the reaction target and the tracking detectors are small, a focused secondary beam is required in order to have a small beam spot on the reaction target. Such a requirement was fulfilled by employing a wedge-shaped aluminum degrader to compensate the momentum deviation (from the reference particle) of the secondary beam. The thickness of the degrader along the optical axis was 5 $\rm{g/cm}^2$ and the wedge angle was 194 mrad. Fig.~\ref{ion-optic_scheme} shows the ion-optical setting of the FRS used in the experiment, which was calculated with the code GICOSY~\cite{Berz1987NIMA,GICOSY}. The colored lines show the trajectories of $^{31}$Ar ions at three different energies, each at five different angles after production by fragmenting a $^{36}$Ar beam on the $^9$Be production target. By combining the ion-optical elements of the FRS (dipole magnets and quadrupole magnets) and the energy loss in the degrader, the optical system TA--F2 was tuned to spatially separate the $^{31}$Ar fragment beam from other fragments and to provide an achromatic image at the middle focal plane F2. The horizontal (X) slits at F1 (not shown in Fig.~\ref{ion-optic_scheme}) and F2 were employed to assist in rejecting the unwanted ions at F2. The second half of the FRS was operated in a dispersive mode and the $^{28}$S ions were transmitted as the centered beam down to F4, where the full Particle IDentification (PID) in A and Z can be performed. The transmission properties of the FRS may be described by the calculated longitudinal momentum ($p$) and angular acceptance. For the section TA--F2, the momentum acceptance ($\Delta p/p$) was limited by closing the slits at F1 to $\Delta p/p=\pm 0.71\%$. The corresponding angular acceptance in the horizontal plane (X plane) was $\pm 14$ mrad, while the angular acceptance in the vertical plane (Y plane) was $\pm 13$ mrad. For a beam between F2 and F4, the momentum acceptance was $\pm 2.8\%$ and the angular acceptance was $\pm 20$ mrad in both X and Y planes.

\begin{figure}[!htbp]
\centerline{\includegraphics[scale=0.32, angle=0]{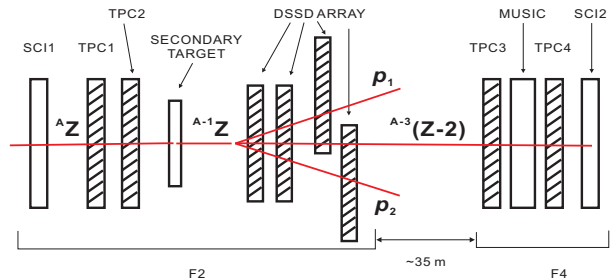}}
\caption{Sketch of the detector setup.
The secondary beam $^AZ$ ($^{31}$Ar or $^{20}$Mg) was tracked by position sensitive detectors TPC1 and TPC2 before impinging on the reaction target. The trajectories of two protons and HI daughter nucleus $^{A-3}(Z-2)$ resulting from the decay of 2\emph{p} precursor $^{A-1}Z$ ($^{30}$Ar or $^{19}$Mg) were measured by the DSSD array. At the focal plane F4, the energy deposition of HI in the detector MUSIC was recorded. The time-of-flight of HI from F2 to F4 ($\sim$35 m) was measured by using the scintillator detectors SCI1 and SCI2.}
\label{DecSetup}
\end{figure}

\begin{figure}[!htbp]
\centerline{\includegraphics[scale=0.45, angle=0]{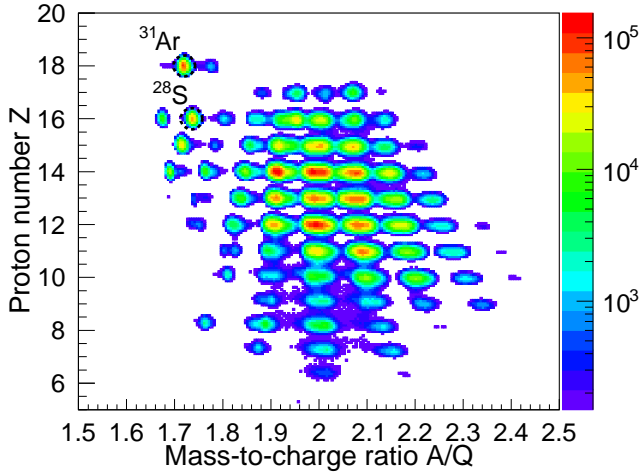}}
\caption{(Color online) Two-dimensional identification plot of $Z$ vs $A/Q$ for the heavy ions detected at F4 during the production measurements with the $^{31}$Ar-$^{28}$S setting. The first half of the FRS was optimized to transport the 620 MeV/u $^{31}$Ar beam and the second half of the FRS was tuned to transmit the $^{28}$S ions.}
\label{ID_Ar}
\end{figure}

The detectors employed in the present experiment are sketched in Fig.~\ref{DecSetup}. The locations of tracking detectors were mainly at the FRS middle focal plane, F2. Two Time-Projection Chambers (TPC1 and TPC2) were used to track the positions of incoming $^{31}$Ar (or $^{20}$Mg) projectiles. A double-sided silicon micro-strip detector (DSSD) array, which consists of four large-area DSSDs~\cite{Stanoiu2008NIMB} was employed to measure hit coordinates of the two protons and the recoil heavy ion ($^{28}$S or $^{17}$Ne) resulting from the in-flight 2\emph{p} decay. The high-precision position measurement by DSSDs allowed us to reconstruct fragment trajectories and to derive the decay vertex together with angular HI-proton and proton-proton correlations. In the second half of the FRS, the heavy ions arriving at the final focal plane of the FRS F4 were unambiguously identified by their magnetic rigidity $B\rho$, time-of-flight (TOF), and energy deposition $\Delta E$. The $B\rho$ of the ion was determined from the FRS magnet setting and the ion's position measured with TPCs.~The TOF for the ion traveling from F2 to F4 was measured by using the scintillator SCI1 at F2 and scintillator SCI2 at F4. Then the ion's velocity ($v$) can be deduced from its TOF. Once the $v$ is obtained, the mass-to-charge ratio ($A/Q$) can be determined by using the following equation
\begin{equation}
\label{eq:AoQ}
\frac{A}{Q}=\frac{B\rho e}{\beta\gamma cu},
\end{equation}
where $e$ is the electron charge, $c$ is the speed of light, $u$ is the atomic mass unit, $\beta$ is the ion's velocity in unit of $c$ ($\beta=v/c$), $\gamma$ is the Lorentz factor ($\gamma=\sqrt{1-\beta^2}$). Given the fact that the energy deposition ($\Delta E$) of the HI in the MUltiple Sampling Ionizing Chamber (MUSIC) is nearly proportional to the square of the ion's charge $Q$, the $Q$ can be calculated form the $\Delta E$ measured by the MUSIC detector. At the high energies used in the present experiment (several hundred MeV/u), most of ions are fully stripped, thus we assume $Q=Z$. Therefore, the HI's proton number $Z$ can be determined from $\Delta E$ measurements. By plotting the distribution of $Z$ versus $A/Q$, the identification of HI can be achieved since each isotope has a unique combination of $Z$ and $A/Q$. Fig.~\ref{ID_Ar} shows a two-dimensional PID plot for the ions which reached F4. In this plot, each nuclide occupies a unique position according to its proton number and mass-to-charge ratio. Therefore, the heavy ion of interest can be identified unambiguously. The ions of interest including $^{28}$S and $^{31}$Ar are well separated from other species and they are highlighted by the circles.

For each identified $^{28}$S nucleus, the coincident protons were identified by registering their impact position in several DSSD's and by requiring a ``straight-line'' trajectory in the analysis. Afterwards, several conditions and thresholds were applied in order to identify the $^{28}$S+\emph{p}+\emph{p} coincidence. The procedure can be divided into three steps. First, the trajectories of protons and of $^{28}$S were reconstructed by using the coordinates measured by DSSDs. Second, the closest approach between proton trajectory and $^{28}$S ion trajectory was measured and tested by checking whether it is a vertex for a $^{28}$S + \emph{p} double coincidence. Third, the difference between the Z coordinates of two vertices (Zdiff) derived from two $^{28}$S + \emph{p} double coincidences was calculated and the Zdiff was tested by another threshold to identify the $^{28}\rm{S}+\emph{p}+\emph{p}$ triple coincidence. If a triple coincidence $^{28}\rm{S}+\emph{p}+\emph{p}$ was identified, an $^{30}$Ar 2\emph{p} decay event was assumed to be found. The detailed description of the search procedure for the 2\emph{p} decay events can be found in Ref.~\cite{Xu2016thesis}.

\section{Investigation of Known Two-Proton Emitter $^{19}$Mg}\label{sec:Reference}

\begin{figure}[!htbp]
\centering
\hspace{-7mm}
\begin{minipage}[b]{0.95\linewidth}
\centering
\includegraphics[scale=0.37, angle=0]{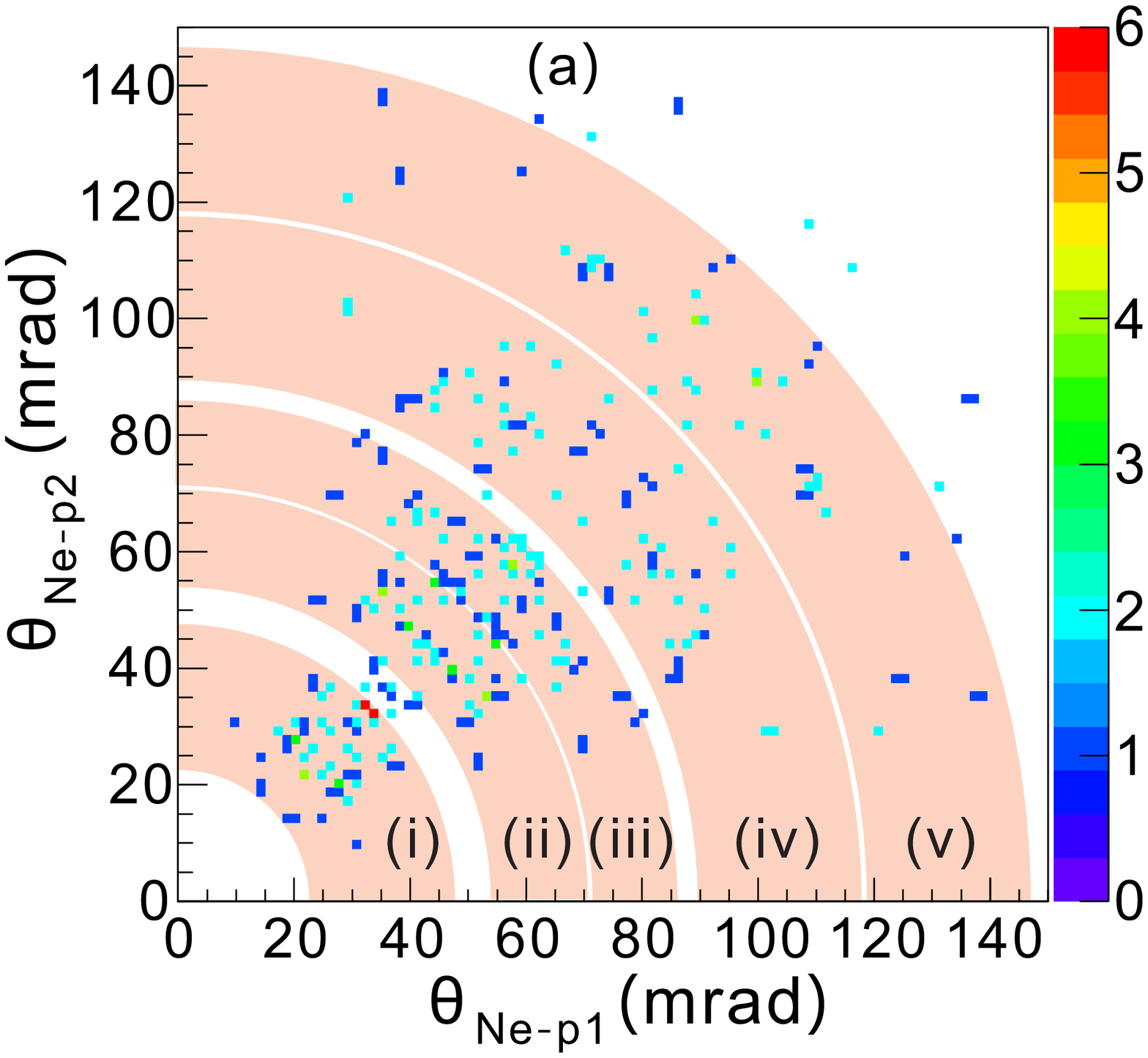}
\end{minipage}
\begin{minipage}[b]{0.95\linewidth}
\centering
\includegraphics[scale=0.4, angle=0]{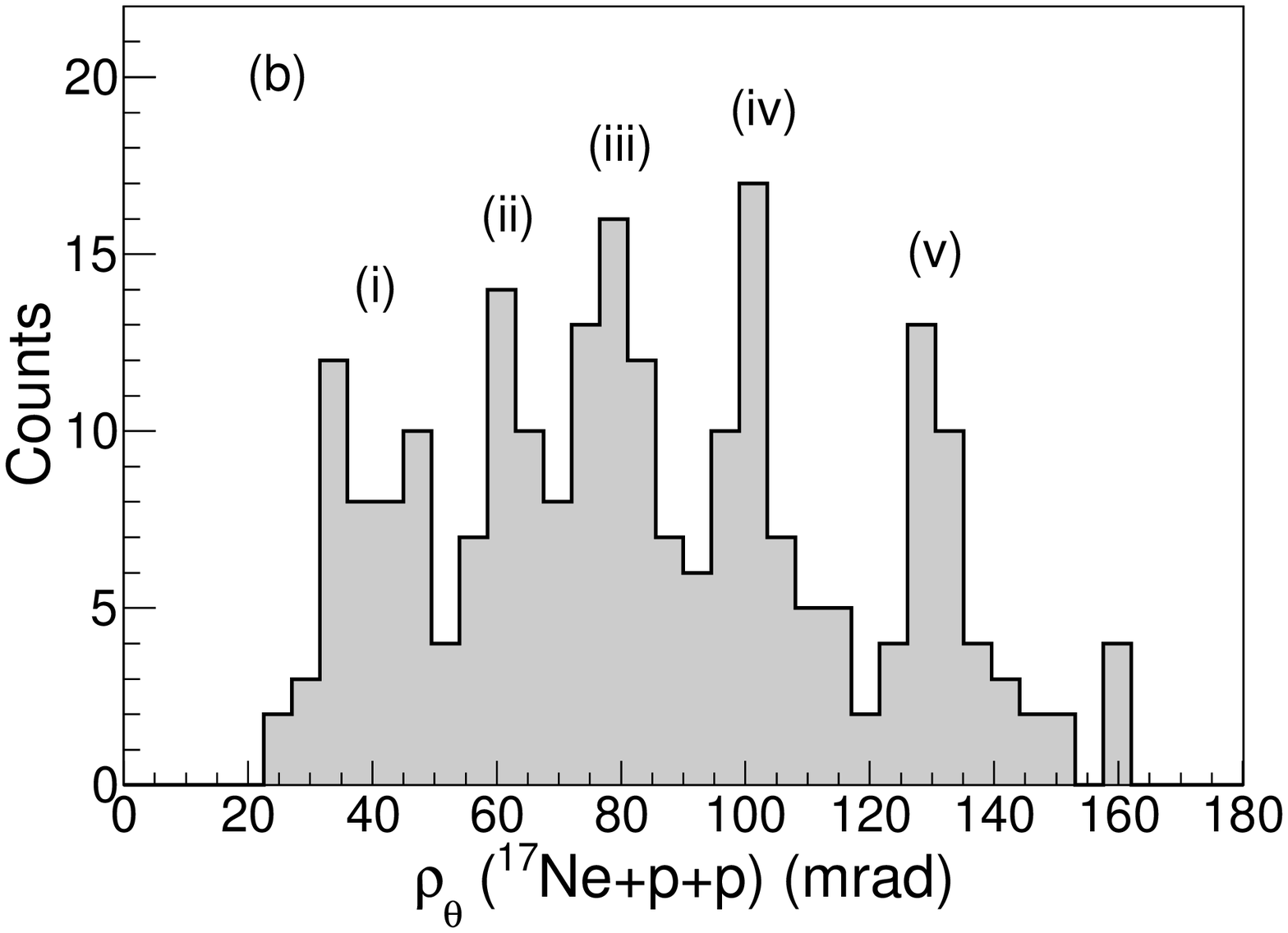}
\end{minipage}
\caption{(Color online) $^{17}$Ne-proton angular correlations derived from the measured $^{17}\rm{Ne}+\emph{p}+\emph{p}$ coincidences. (a) Angular correlations $\theta_{\rm{Ne}\text{-}p1}$-$\theta_{\rm{Ne}\text{-}p2}$. (b) Measured $\rho_\theta$ spectrum for 2\emph{p} decays of $^{19}$Mg. The peak in panel (b) and the arc in panel (a) labeled with the same Roman numeral correspond to each other.}
\label{theta_rho_Mg}
\end{figure}

For calibration purposes, the previously-known 2\emph{p} radioactive nuclei $^{19}$Mg were also produced by a $1n$ knockout reaction from $^{20}$Mg ions obtained by fragmenting a 685 MeV/u $^{36}$Ar beam. The 2\emph{p} decay properties of $^{19}$Mg were remeasured.
By following the same procedure applied in previous studies of $^{19}$Mg \cite{Mukha2007PRL,Mukha2010PRC,Mukha2012PRC}, the decay properties of the precursors $^{19}$Mg and $^{30}$Ar were investigated on the basis of angular correlations between the HI daughter nucleus and the protons. In this section, the angular $^{17}$Ne-proton correlations obtained from 2\emph{p} decays of $^{19}$Mg are described. Based on the measured trajectories of $^{17}$Ne and two protons which were emitted by the 2\emph{p} decay of $^{19}$Mg, the angle between the $^{17}$Ne and proton's trajectories ($\theta_{\rm{Ne}\text{-}p}$) as well as the angle between both protons' trajectories ($\theta_{p\text{-}p}$) can be obtained. The corresponding $^{17}$Ne-\emph{p} angular correlations were reconstructed for all $^{17}\rm{Ne}+\emph{p}+\emph{p}$ coincidences. Fig.~\ref{theta_rho_Mg}(a) shows the scatter plot ($\theta_{\rm{Ne}\text{-}p1}$, $\theta_{\rm{Ne}\text{-}p2}$) for the measured angles between $^{17}$Ne and both protons. Since the two protons cannot be distinguished, the distribution is symmetrized with respect to proton permutations, and proton indexes are given for illustration purpose only. In this angular correlation plot, there are several intensity enhancements which provide the information on the 2\emph{p} states in $^{19}$Mg and 1\emph{p} resonances in $^{18}$Na.

In order to better reveal the decay properties from the measured $^{17}$Ne-proton angular correlations shown in Fig.\ \ref{theta_rho_Mg} (a), one may use the fact that the two protons emitted by one state of $^{19}$Mg share the total decay energy. Thus $\theta_{\rm{Ne}\text{-}p}$ correlations from 2\emph{p} decays of the same narrow state are accumulated along the arc with the radius
\[
\rho_\theta(^{17}\text{Ne}+p+p)=\sqrt{\theta^{2}_{\text{Ne}\text{-}p_1}+\theta^{2}_{\text{Ne}\text{-}p_2}}=\text{const} \, .
\]
Since $\rho_\theta$ is related to the energy sum of both emitted protons and therefore to the $Q_{2p}$ of the parent state by the relation $Q_{2p}\sim\rho_{\theta}^2$~\cite{Mukha2012PRC}, one can obtain the indication of the parent state and its 2\emph{p}-decay energy by studying the distribution of $\rho_\theta$. The $\rho_\theta$ spectrum has a few peaks which allow us to select specific excitation-energy regions for the investigation. In the present study, the $\rho_\theta$ distribution measured for $^{19}$Mg 2\emph{p} decays is displayed in Fig.\ \ref{theta_rho_Mg} (b). Several well-separated intense peaks, which indicate the 2\emph{p} decays of various states in $^{19}$Mg, are clearly seen and labeled by Roman numerals. The peak in Fig.\ \ref{theta_rho_Mg} (b) corresponds to the arc in Fig.\ \ref{theta_rho_Mg} (a) labeled with the same Roman numeral. By gating on a particular $\rho_\theta$ peak, the decay events from a certain $^{19}$Mg state can be selected. In the following, the states observed in $^{19}$Mg will be investigated by comparing the measured $^{17}$Ne-\emph{p} angular correlations with the Monte Carlo (MC) simulations of the detector response.

\subsection{Reference case: 1\emph{p} and 2\emph{p} decays of known states in $^{18}$Na and $^{19}$Mg}
\label{subsec:Mg_Na_known}

\begin{figure*}[!htbp]
\centering
\begin{minipage}[b]{0.45\linewidth}
\centering
\includegraphics[scale=0.35, angle=0]{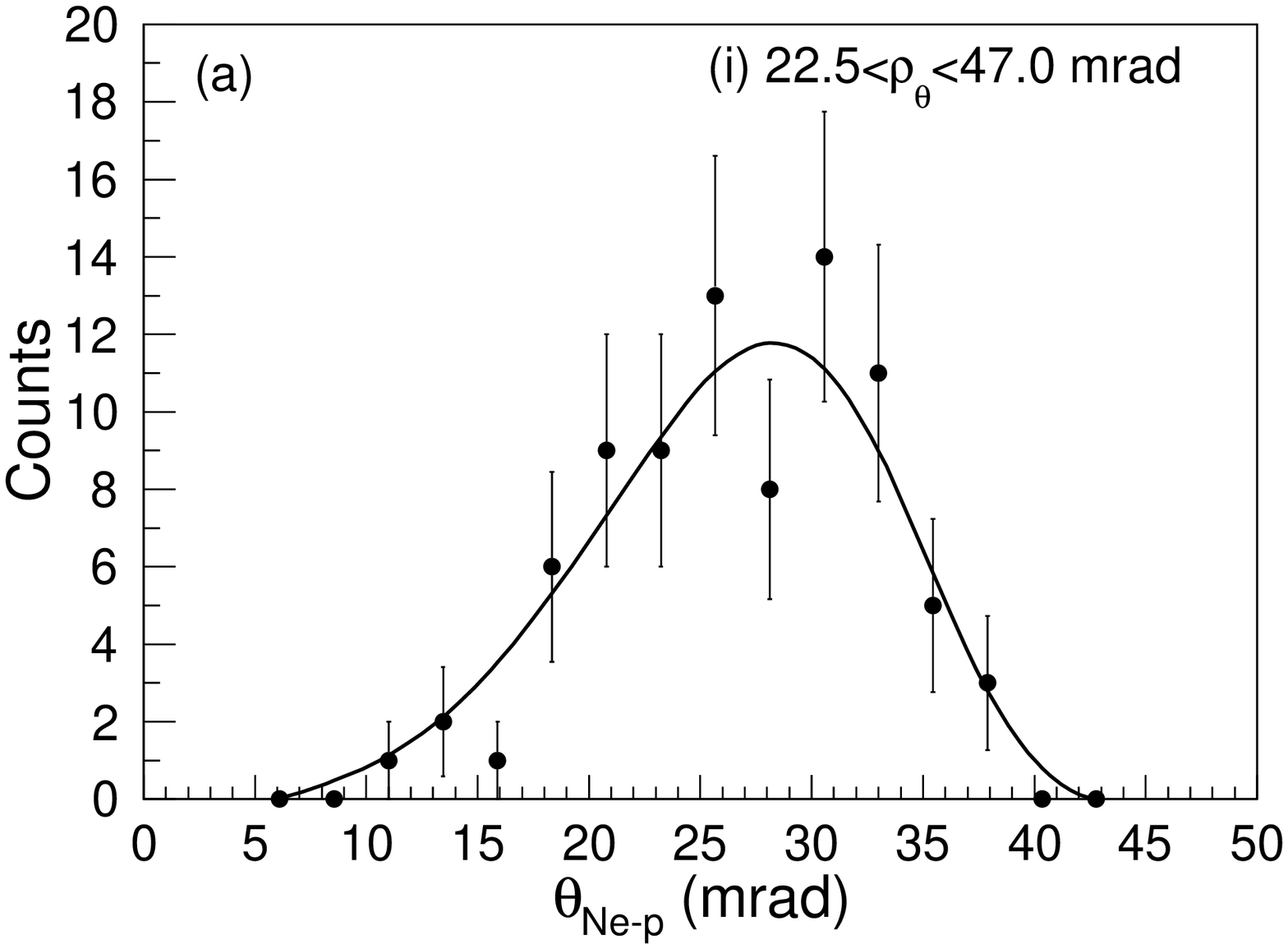}
\end{minipage}
\begin{minipage}[b]{0.45\linewidth}
\centering
\includegraphics[scale=0.35, angle=0]{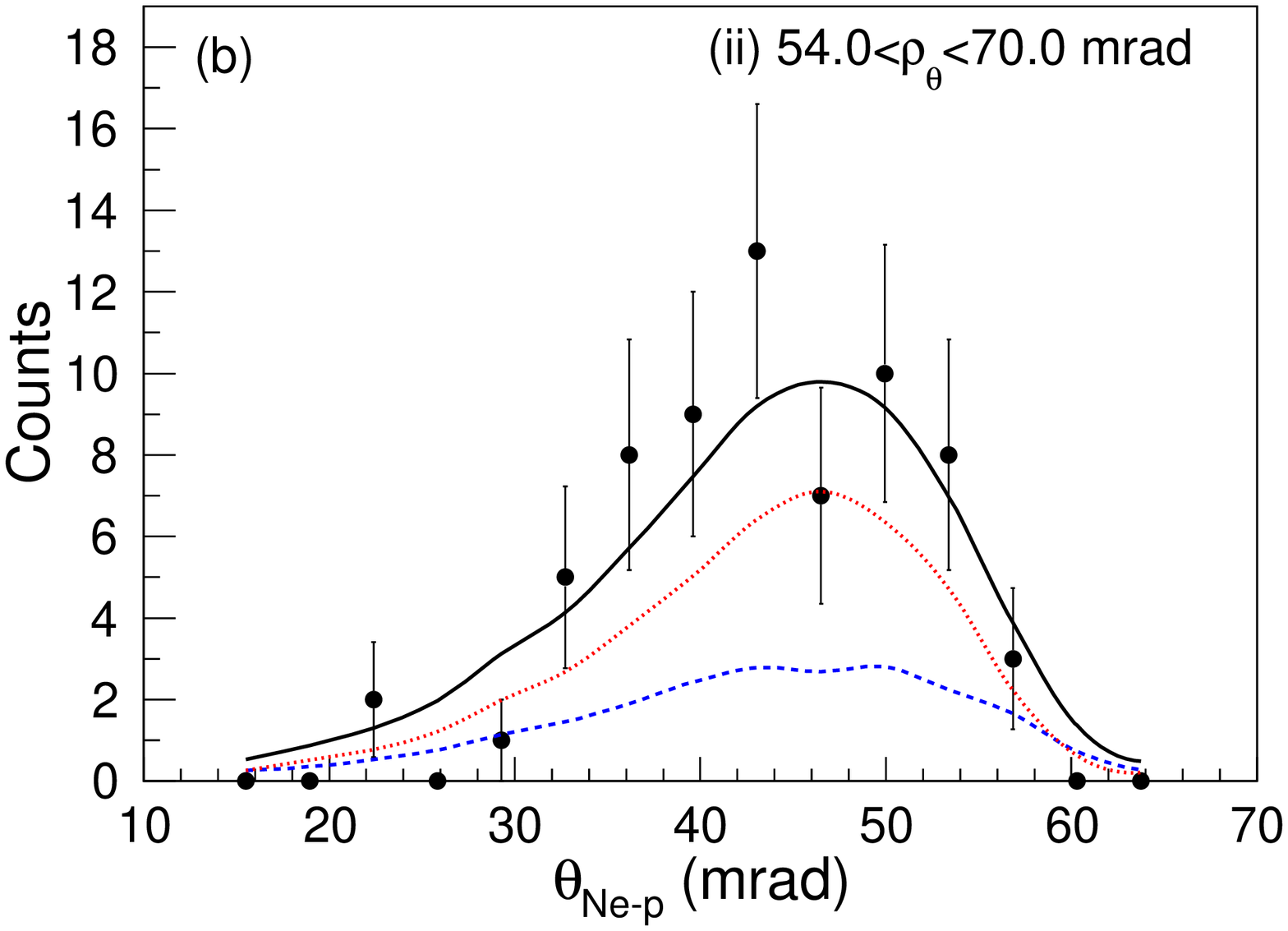}
\end{minipage}
\centering
\begin{minipage}[b]{0.45\linewidth}
\centering
\includegraphics[scale=0.35, angle=0]{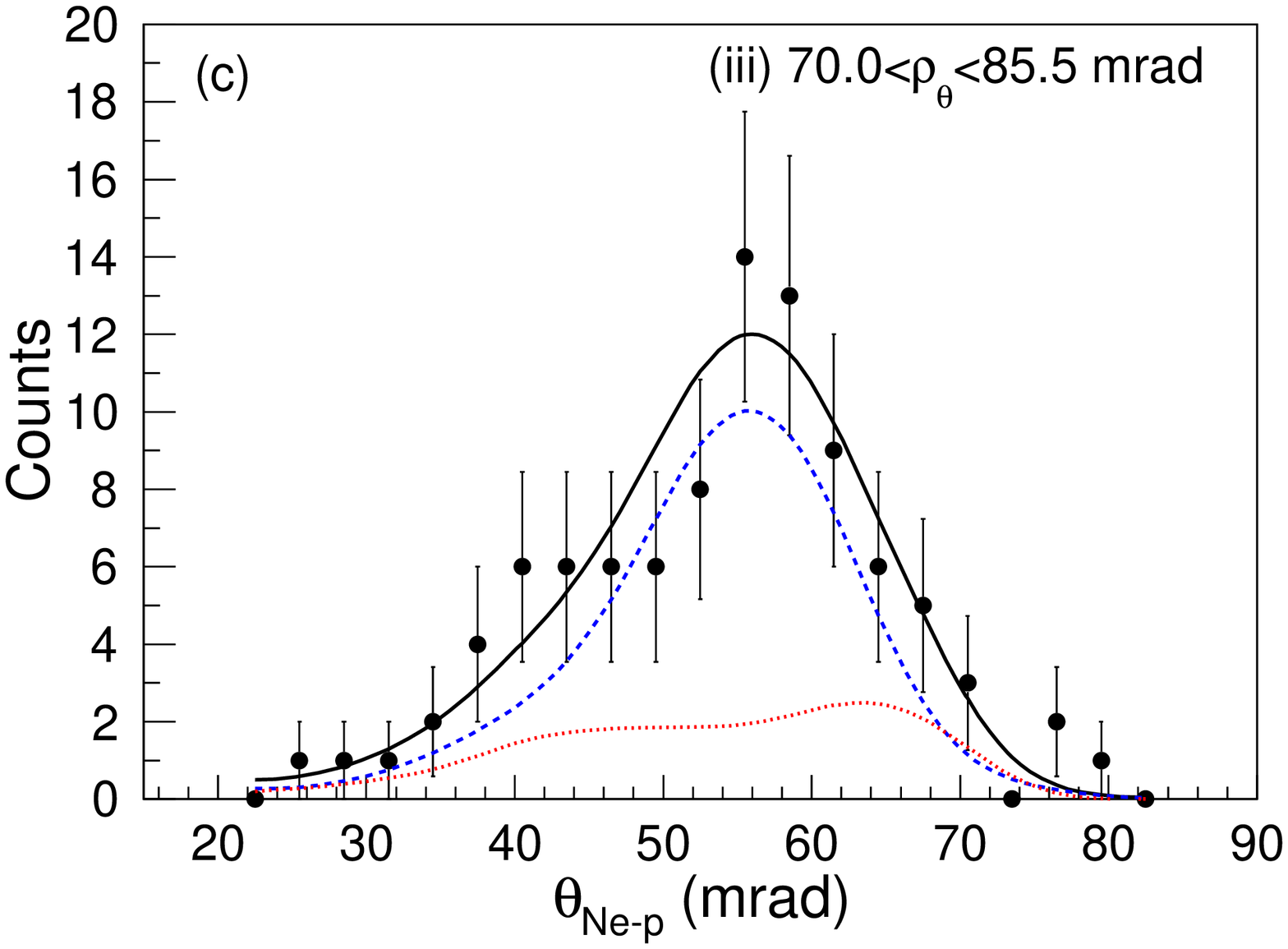}
\end{minipage}
\begin{minipage}[b]{0.45\linewidth}
\centering
\includegraphics[scale=0.35, angle=0]{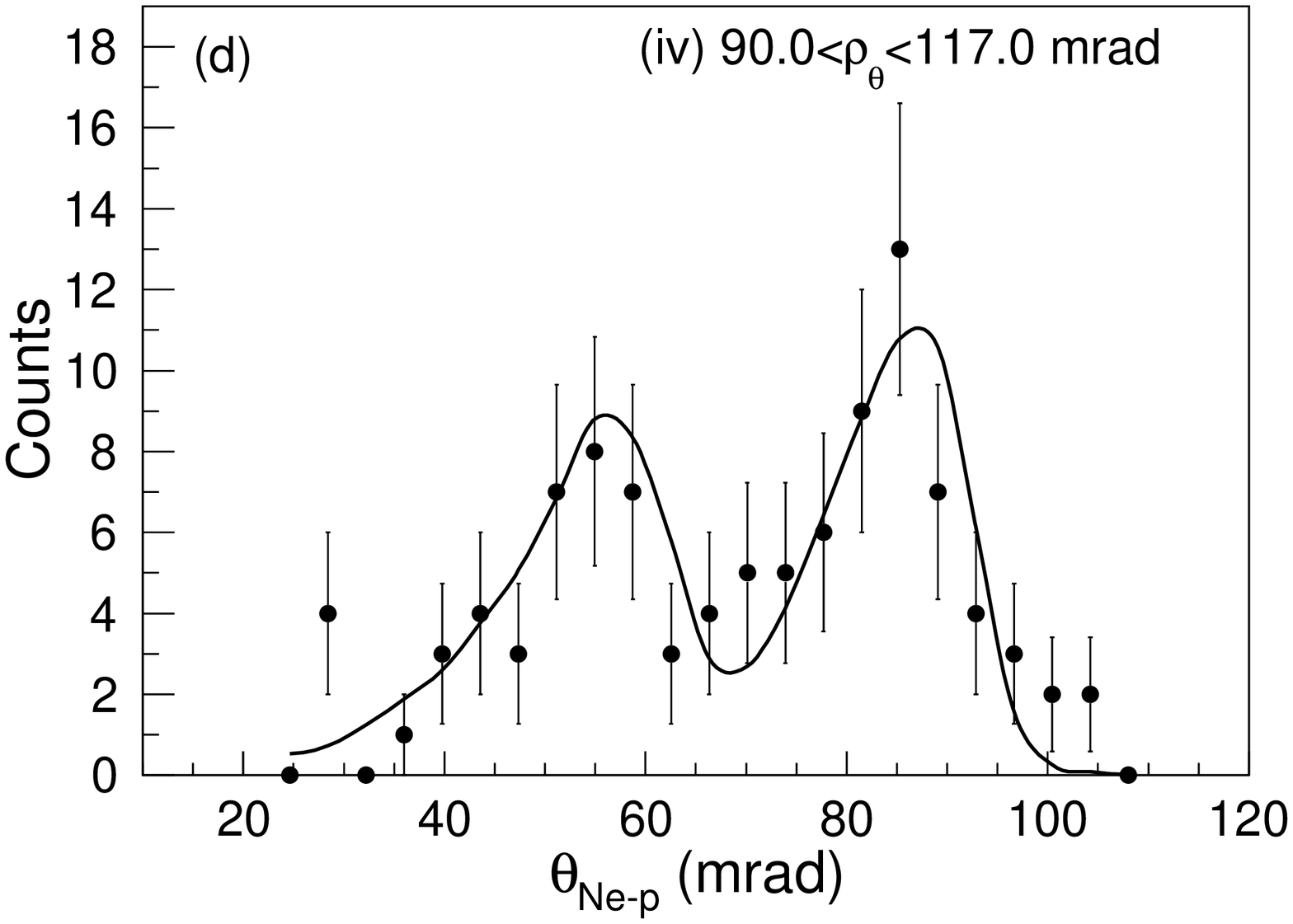}
\end{minipage}
\begin{minipage}[b]{0.45\linewidth}
\centering
\includegraphics[scale=0.35, angle=0]{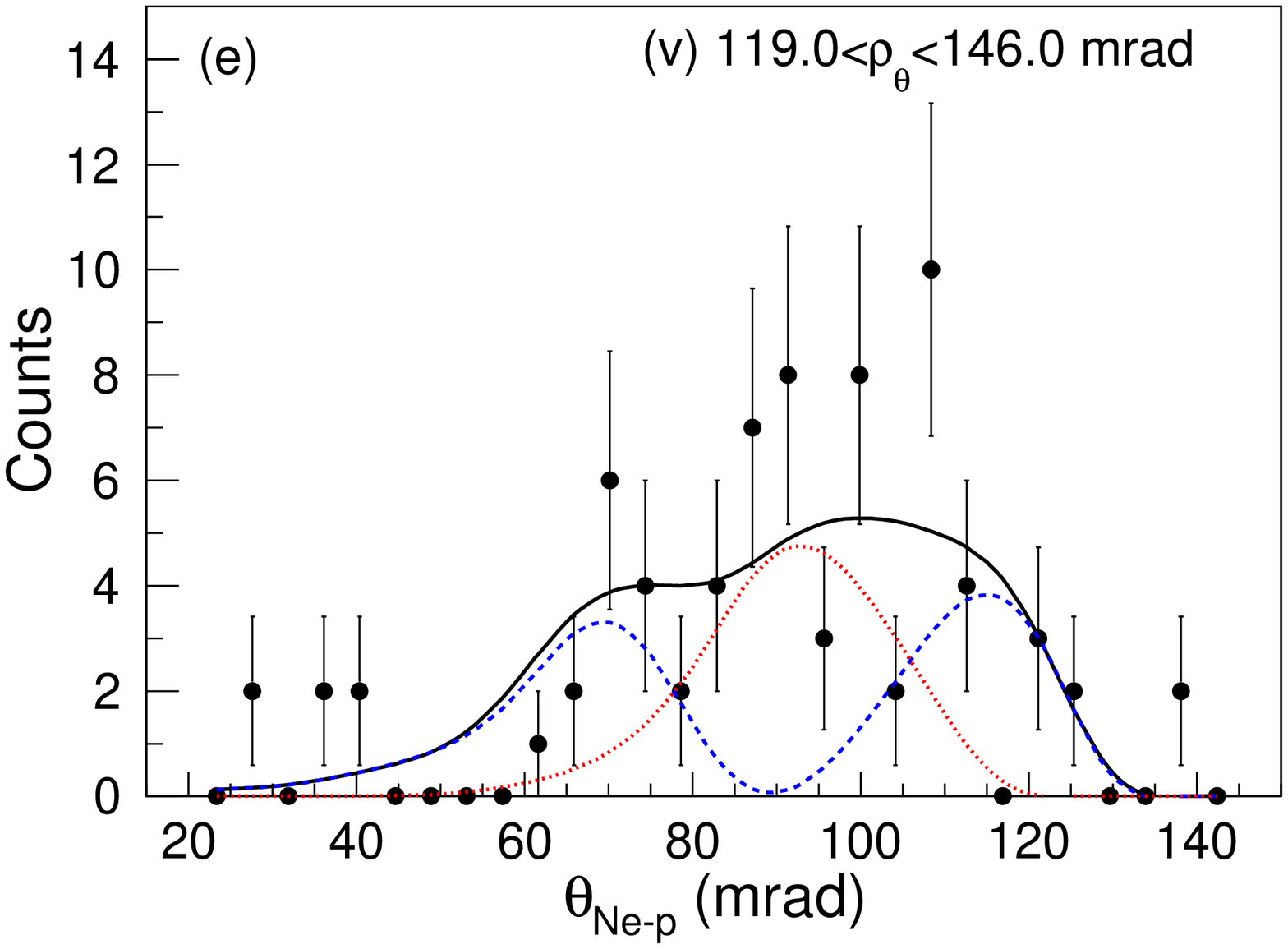}
\end{minipage}
\caption{(Color online) Measured $^{17}$Ne-\emph{p} angular correlations (full circles with statistical errors) derived from the 2\emph{p} decays of known $^{19}$Mg states. (a) Measured $^{17}$Ne-\emph{p} angular correlations derived from the 2\emph{p} decay of $^{19}$Mg g.s.~gated by (i), $22.5 < \rho_\theta < 47.0$ mrad. The solid curve represents the corresponding MC simulation of the detector response to the simultaneous 2\emph{p} decay of the $^{19}$Mg g.s.~with $Q_{2p} = 0.87$ MeV. (b) The 2\emph{p} decay of the excited state gated by (ii), $54.0<\rho_\theta<70.0$ mrad. The solid curve displays the simulation of the sequential 2\emph{p} decay of $^{19}$Mg state at 2.5 MeV via $^{18}$Na states at 1.23 MeV (dotted curve) and 1.55 MeV (dashed curve). (c) The 2\emph{p} decays gated by (iii), $70.0<\rho_\theta<85.5$ mrad. The solid curve is the simulation of the sequential 2\emph{p} decay of $^{19}$Mg state at 3.2 MeV via the 1.55 MeV (dashed curve) and 2.084 MeV (dotted curve) levels in $^{18}$Na. (d) The 2\emph{p} decays gated by (iv), $90.0<\rho_\theta<117.0$ mrad. The result of the simulation to the sequential 2\emph{p} emission of $^{19}$Mg state at 5.1 MeV via 1.55 MeV state of $^{18}$Na state is depicted by the solid curve. (e) The 2\emph{p} decay of a new excited state in $^{19}$Mg gated by (v), $119.0<\rho_\theta<146.0$ mrad. The dashed and dotted curves are the $\theta_{\rm{Ne}\text{-}p}$ distributions obtained by simulations of sequential proton emission of $^{19}$Mg state at 8.9 MeV via two unknown observed $^{18}$Na states at 2.5 MeV and 4.0 MeV, respectively. The solid curve shows the summed fit.}
\label{ES_Mg}
\end{figure*}

By comparing the $\theta_{\rm{Ne}\text{-}p}$ angular correlations [Fig.~\ref{theta_rho_Mg}(a)] with those obtained in the previous experiment (see Figure 2(c) in Ref.~\cite{Mukha2012PRC}), it was found that several known states of $^{19}$Mg including its g.s.~and several low-lying excited states were observed in the present experiment. They are shown by the peaks and arcs (i), (ii), (iii),
and (iv) in Fig.~\ref{theta_rho_Mg}. In order to quantitatively interpret the $^{17}$Ne-\emph{p} angular correlations obtained from the decays of these known states, MC simulations were performed by assuming the simultaneous 2\emph{p} decay of the $^{19}$Mg g.s.~and the sequential 2\emph{p} decay of $^{19}$Mg excited states via low-lying $^{18}$Na states. The simulated $\theta_{\rm{Ne}\text{-}p}$ distributions were compared with the data obtained by choosing events with the $\rho_\theta$ gates (i), (ii), (iii), and (iv) indicated in Fig.~\ref{theta_rho_Mg} (b). The corresponding results are shown in the panels (a), (b), (c) and (d) of Fig.~\ref{ES_Mg}, respectively. One can see that the simulations reproduce the data well. The deduced 2\emph{p}-decay energy ($Q_{2p}$) of the g.s.~shown in (i) is $0.87^{+0.24}_{-0.07}$ MeV, which is consistent with the previous result of 0.76(6) MeV~\cite{Mukha2012PRC}. The deduced $Q_{2p}$ values of the excited state (ii) and excited state (iv) are $2.5^{+0.8}_{-0.4}$ MeV and $5.1^{+0.3}_{-0.3}$ MeV, respectively, which agree within the errors with the previous data on the respective states at 2.14(23) MeV and 5.5(2) MeV~\cite{Mukha2012PRC}. The determined $Q_{2\emph{p}}$ for the broad peak (iii) is $3.2^{+1.2}_{-1.0}$ MeV, which matches the previously-measured states at 2.9(2) and 3.6(2) MeV. However, these two states cannot be resolved in the present experiment.

\subsection{Hints to so far unknown 1\emph{p}- and 2\emph{p}- unbound states in $^{18}$Na and $^{19}$Mg}
\label{subsec:Mg_Na_unknown}

In Fig.~\ref{theta_rho_Mg}(b), besides the known excited states of $^{19}$Mg shown by the peaks (ii), (iii), and (iv), evidence on a new excited state of $^{19}$Mg is displayed by the peak (v) which is located around $\rho_\theta=130$ mrad. The corresponding $\theta_{\rm{Ne}\text{-}p1}$ versus $\theta_{\rm{Ne}\text{-}p2}$ distribution is shown by the arc (v) in Fig.~\ref{theta_rho_Mg}(a). One can see that most events fall into several clusters which indicate sequential emission of protons from one excited state of $^{19}$Mg via intermediate resonances of $^{18}$Na. It is worth mentioning that the hints to sequential proton emission from such an unknown excited state of $^{19}$Mg can be also found in the experimental spectrum obtained from a previous study of $^{19}$Mg (see Fig. 2(c) of~\cite{Mukha2012PRC}). Despite the low counts, the peak (v) and multiple-cluster structure in the corresponding arc (v) are quite evident, which may be attributed to a different detection scheme being applied thus leading to a better signal to noise ratio in the present experiment. The angular $\theta_{\rm{Ne}\text{-}p}$ spectrum obtained by imposing the arc gate (v) ($119.0<\rho_\theta<146.0$ mrad) is shown by the black dots in Fig.~\ref{ES_Mg}(e). Such a multiple peak structure cannot be described by a sequential 2\emph{p} decay via any previously-known $^{18}$Na state because the characteristic $\theta_{\rm{Ne}\text{-}p}$ pattern generated from 1\emph{p} decay of the known state in $^{18}$Na does not fit any peak shown in Fig.~\ref{ES_Mg}(e). In order to interpret such experimental $\theta_{\rm{Ne}\text{-}p}$ spectrum, the existence of two new $^{18}$Na levels has to be assumed.

\begin{figure}[!htbp]
\centerline{\includegraphics[scale=0.33, angle=0]{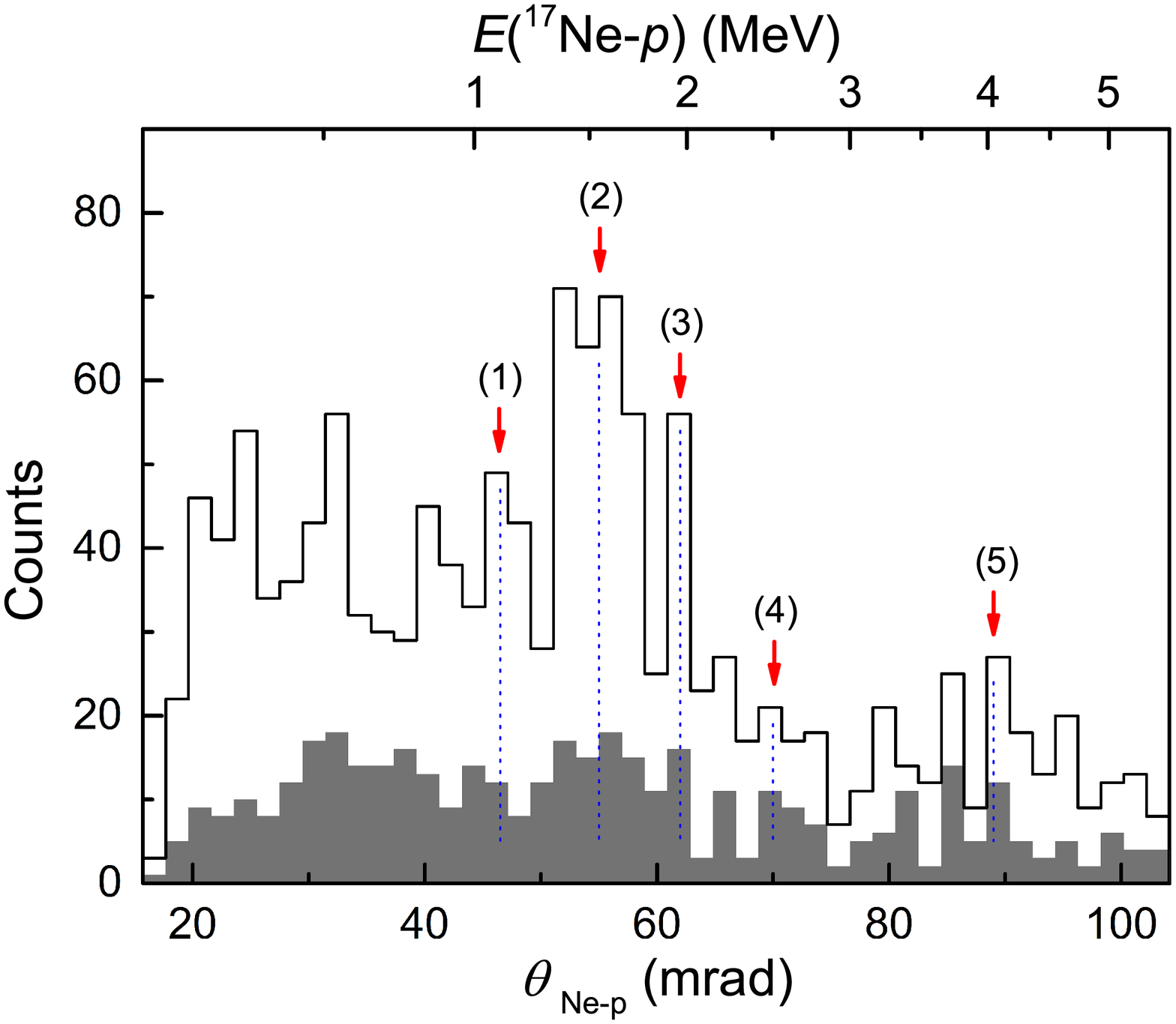}}
\caption{$\theta_{\rm{Ne}\text{-}p}$ distribution derived from the measured $^{17}\rm{Ne}+\emph{p}$ coincidences (unfilled histogram)
and that deduced from the $^{17}\rm{Ne}+\emph{p}+\emph{p}$ coincidences (grey-filled histogram). The blue dashed lines together with
red arrows indicate the peaks which appear in both histograms and these peaks suggest the $^{18}$Na resonances. Previously-known states
of $^{18}$Na are shown by peaks (1), (2), (3), while the peaks (4) and (5) suggest two new resonances in $^{18}$Na.
Corresponding 1\emph{p}-decay energies are shown in the upper axis in MeV. See text for details.}
\label{Theta_Mg_23}
\end{figure}

The hints of two new states expected for $^{18}$Na can be found in Fig.~\ref{Theta_Mg_23}, which displays the comparison of angular $\theta_{\rm{Ne}\text{-}p}$ correlations obtained from the measured $^{17}\rm{Ne}+\emph{p}$ and $^{17}\rm{Ne}+\emph{p}+\emph{p}$ coincidences in the present experiment. In the former case, parent $^{18}$Na states may be populated via several possible reactions on $^{20}$Mg, while the latter distribution is presumably due to the 2\emph{p} emission from $^{19}$Mg states. Five peaks (1-5) which coexist in both histograms suggest the states of $^{18}$Na. According to the previous knowledge on the angular correlations obtained from the decays of known $^{18}$Na states~\cite{Mukha2012PRC}, the peaks 1-3 correspond to the the $^{18}$Na states at 1.23 MeV, 1.55 MeV, and 2.084 MeV, respectively. The peaks 4 and 5 provide indications of two unknown excited states in $^{18}$Na, which are located at 2.5 MeV and 4.0 MeV above the 1\emph{p} threshold, respectively.

The two above-discussed new states in $^{18}$Na provide a possible explanation for the observed $\theta_{\rm{Ne}\text{-}p}$ distribution shown in Fig.~\ref{ES_Mg}(e), i.e., such a $^{17}$Ne-proton angular correlations may originate from the decays of a previously unknown excited state in $^{19}$Mg by sequential emission of protons via the above-mentioned two $^{18}$Na excited states To verify such a tentative assignment, MC simulations were performed. By varying the decay energies and lifetimes of $^{19}$Mg state and $^{18}$Na levels, we found that the simulation of sequential emission of protons from the $^{19}$Mg excited state at $8.9^{+0.8}_{-0.7}$ MeV via the excited states of $^{18}$Na at $2.5^{+0.7}_{-0.3}$ MeV and $4.0^{+1.5}_{-0.6}$ MeV can reproduce the data. The corresponding two components are displayed by the dashed and dotted curves in Fig.~\ref{ES_Mg}(e), respectively. The summed fit generally agrees with the data. In particular, the multiple-peak structure of the experimental pattern is reasonably described. The energy level of $^{18}$Na around 2.5 MeV has been predicted in a theoretical work~\cite{Fortune2007PRC}. Given the fact that the limited amount of $^{19}$Mg 2\emph{p} decay events identified in the present experiment provides only hints of a new $^{19}$Mg excited state and two new $^{18}$Na excited states, future experiments with improved conditions (e.g., better statistics) are desirable.

\section{Spectroscopy of States Observed in $^{30}$Ar and $^{29}$Cl}\label{sec:Ar_ex}

\subsection{$^{28}$S-proton angular correlations}\label{subsec:Correlations}

\begin{figure}[!htbp]
\centering
\hspace{-7mm}
\begin{minipage}[b]{0.95\linewidth}
\centering
\includegraphics[scale=0.37, angle=0]{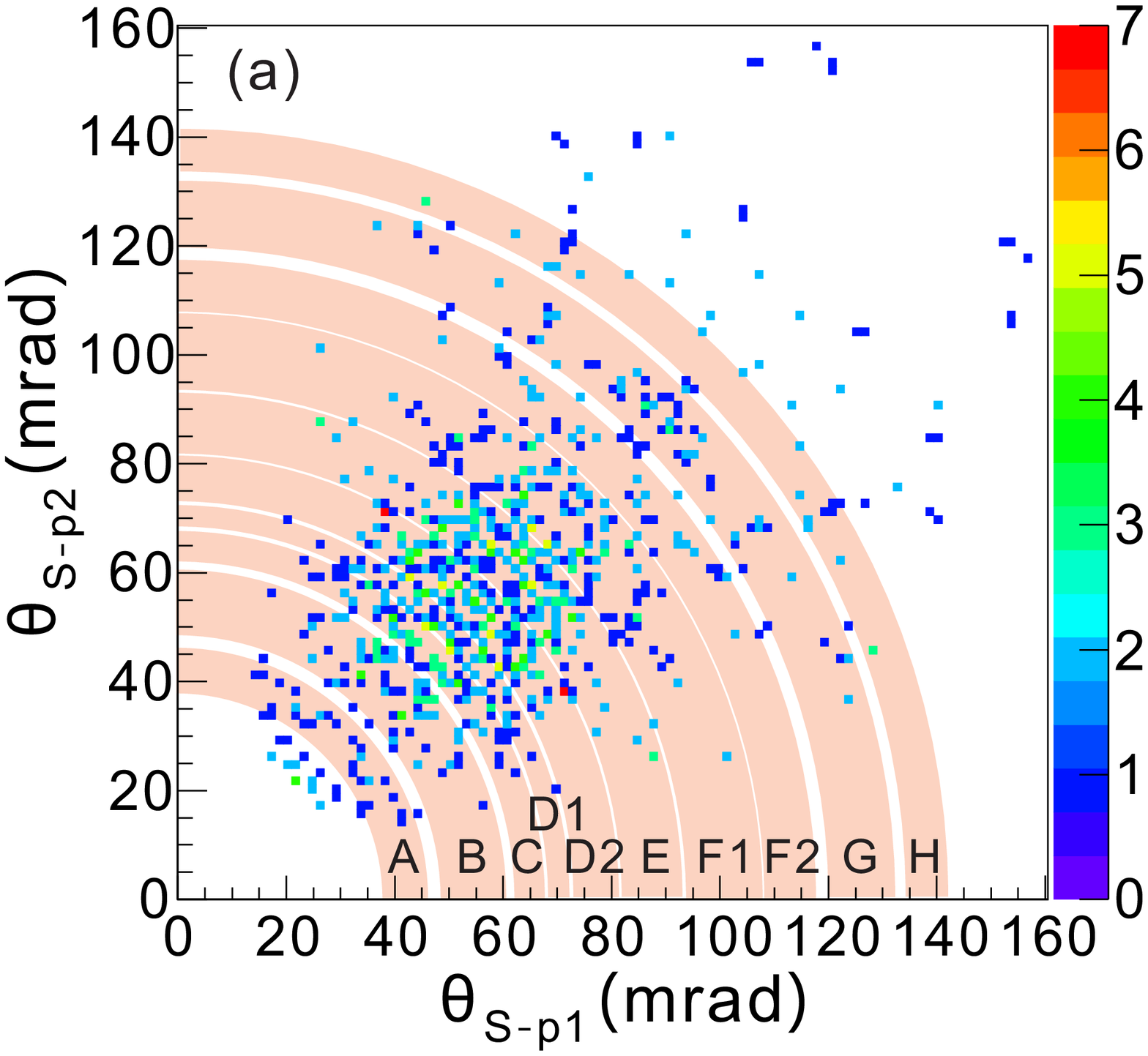}
\end{minipage}
\begin{minipage}[b]{0.95\linewidth}
\centering
\includegraphics[scale=0.35, angle=0]{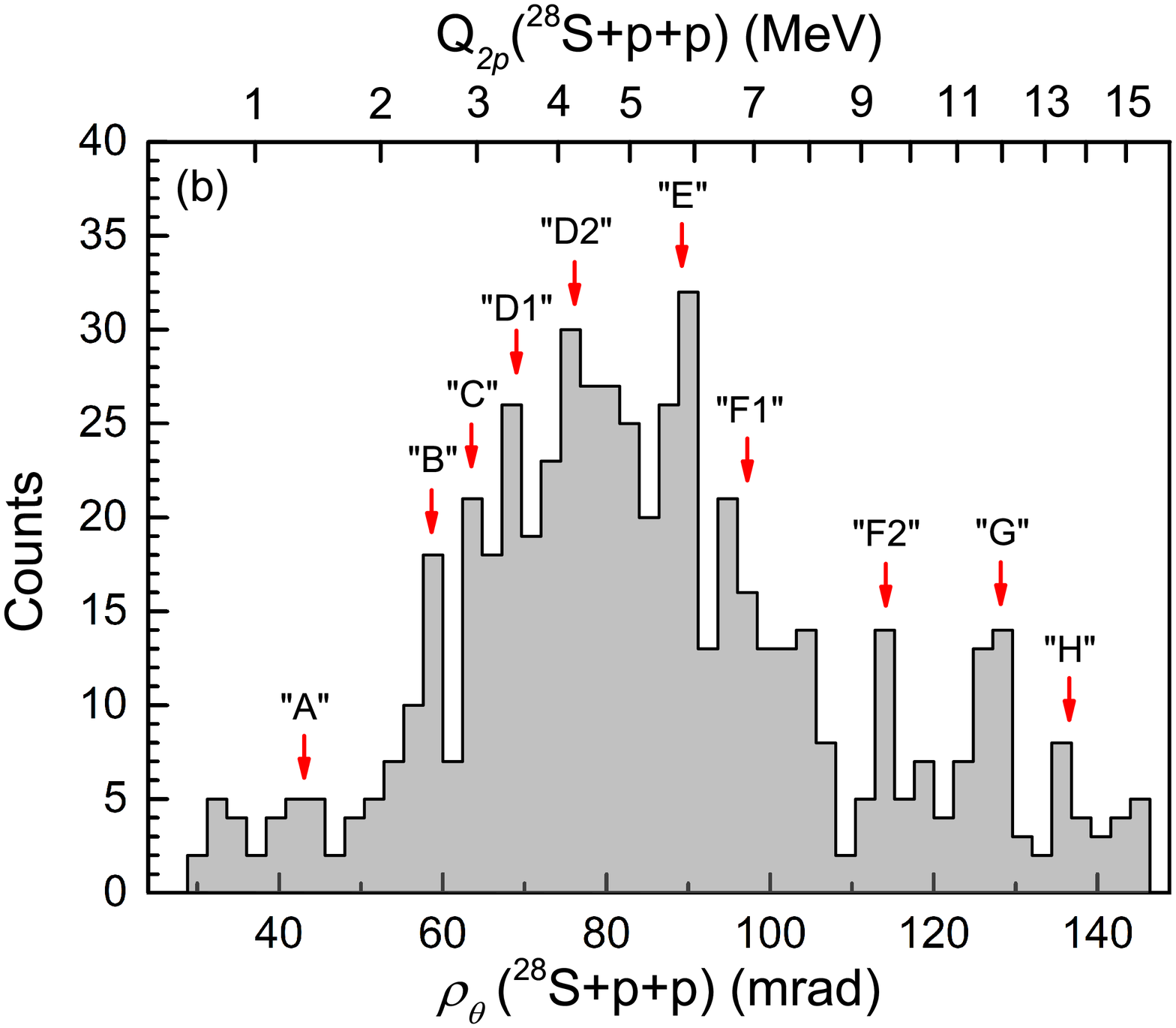}
\end{minipage}
\caption{(Color online) $^{28}$S-proton angular correlations derived from measured $^{28}\rm{S}+\emph{p}+\emph{p}$ coincidences. (a) $\theta_{\rm{S}\text{-}p1}$ versus $\theta_{\rm{S}\text{-}p2}$ distribution. (b) The corresponding $\rho_\theta$ spectrum. The peaks and respective arcs labeled with ``A-H'' suggest the states of $^{30}$Ar. Corresponding 2\emph{p}-decay energies are displayed in the upper axis in MeV. See text for details.}
\label{Theta_rho_Ar}
\end{figure}

As described in Sec.\ \ref{sec:Experiment}, the decays of $^{30}$Ar were identified by tracking the coincident $^{28}\rm{S}+\emph{p}+\emph{p}$ trajectories. Following a similar procedure to that conducted for 2\emph{p} decays of $^{19}$Mg, we measured the angles between the decay products of $^{30}$Ar (i.e., $\theta_{\rm{S}\text{-}p}$ and $\theta_{p1\text{-}p2}$) and then reconstructed the $^{28}$S-proton angular correlations as well as the decay vertices. The scatter plot of $\theta_{\rm{S}\text{-}p1}$ versus $\theta_{\rm{S}\text{-}p2}$ for all identified $^{28}\rm{S}+\emph{p}+\emph{p}$ coincidences is shown in Fig.~\ref{Theta_rho_Ar}(a)$.$ Here proton indexes are given for illustration purposes only. Several intensity enhancements can be observed in this angular correlation plot, and they indicate on the 2\emph{p} states in $^{30}$Ar and 1\emph{p} resonances in $^{29}$Cl. The arcs labeled ``A-H'' in Fig.~\ref{Theta_rho_Ar}(a) correspond to peaks in the $\rho_\theta$ spectrum shown in Fig.~\ref{Theta_rho_Ar}(b), where $\rho_\theta=\sqrt{\theta^{2}_{\rm{S}\text{-}p1}+\theta^{2}_{\rm{S}\text{-}p2}}.$ As demonstrated in the previous chapter, the $\rho_\theta$ distribution is helpful in order to identify the states of $^{30}$Ar and to discriminate transitions of interest. In Fig.~\ref{Theta_rho_Ar}(b), the peaks labeled ``A-H'' suggest several states in $^{30}$Ar, and the corresponding arcs in Fig.~\ref{Theta_rho_Ar}(a) illustrate the $^{28}$S-proton angular correlation patterns. These arcs and peaks demonstrate the first observation of 2\emph{p} decays from several states of the nucleus $^{30}$Ar. In order to deduce the nuclear structure information on these states and investigate their decay properties, careful analysis of $\theta_{\rm{S}\text{-}p}$ patterns and detailed theoretical calculations as well as MC simulations were performed.

\subsection{$\theta_{\rm{S}\text{-}p}$ distributions of states observed in $^{30}$Ar}
\label{subsec:theta_S_p}

\begin{figure*}[!htbp]
\centering
\hspace{-3em}
\begin{minipage}[b]{0.45\linewidth}
\centering
\includegraphics[scale=0.5,angle=0]{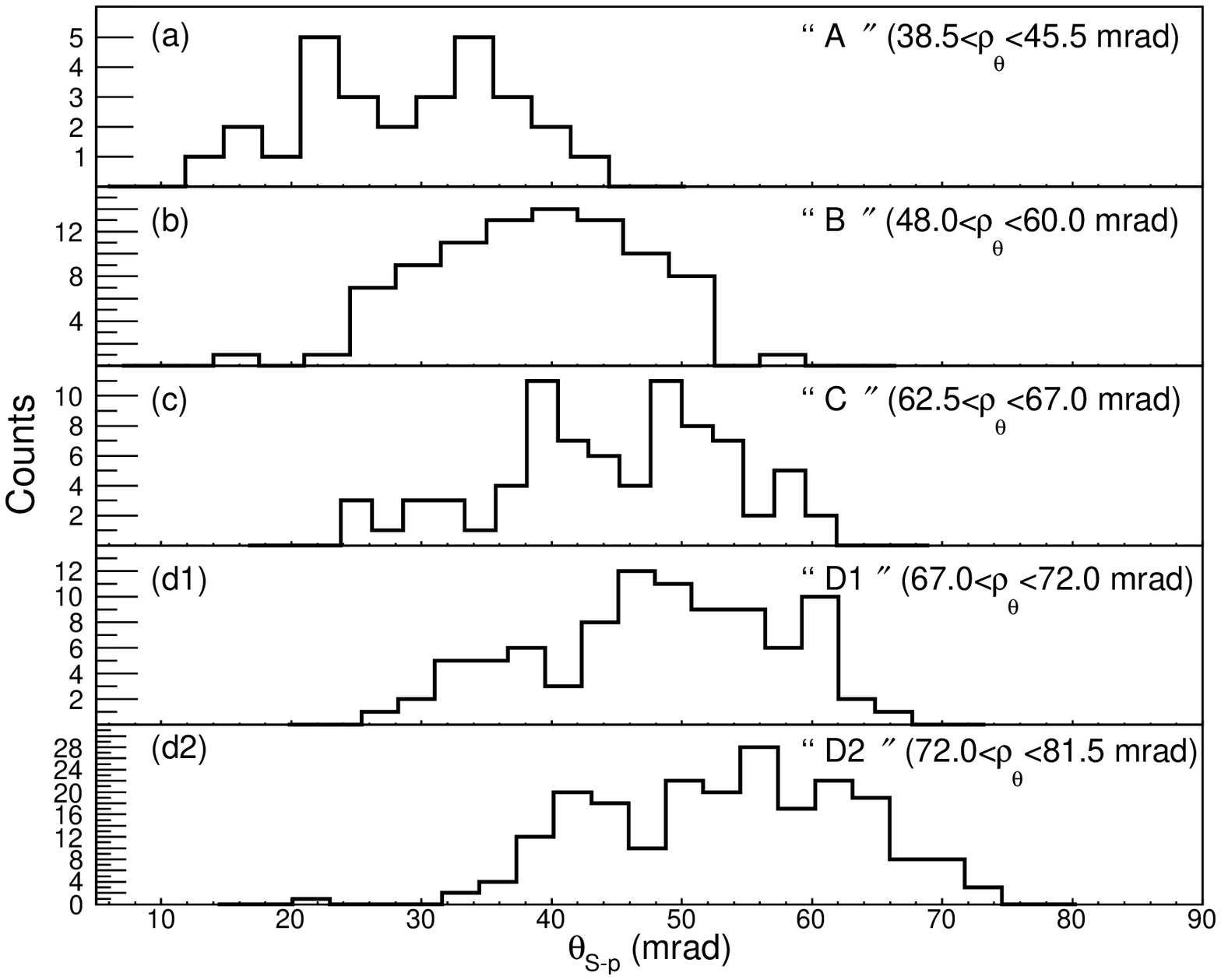}
\end{minipage}
\hspace{3em}
\begin{minipage}[b]{0.45\linewidth}
\centering
\includegraphics[scale=0.5,angle=0]{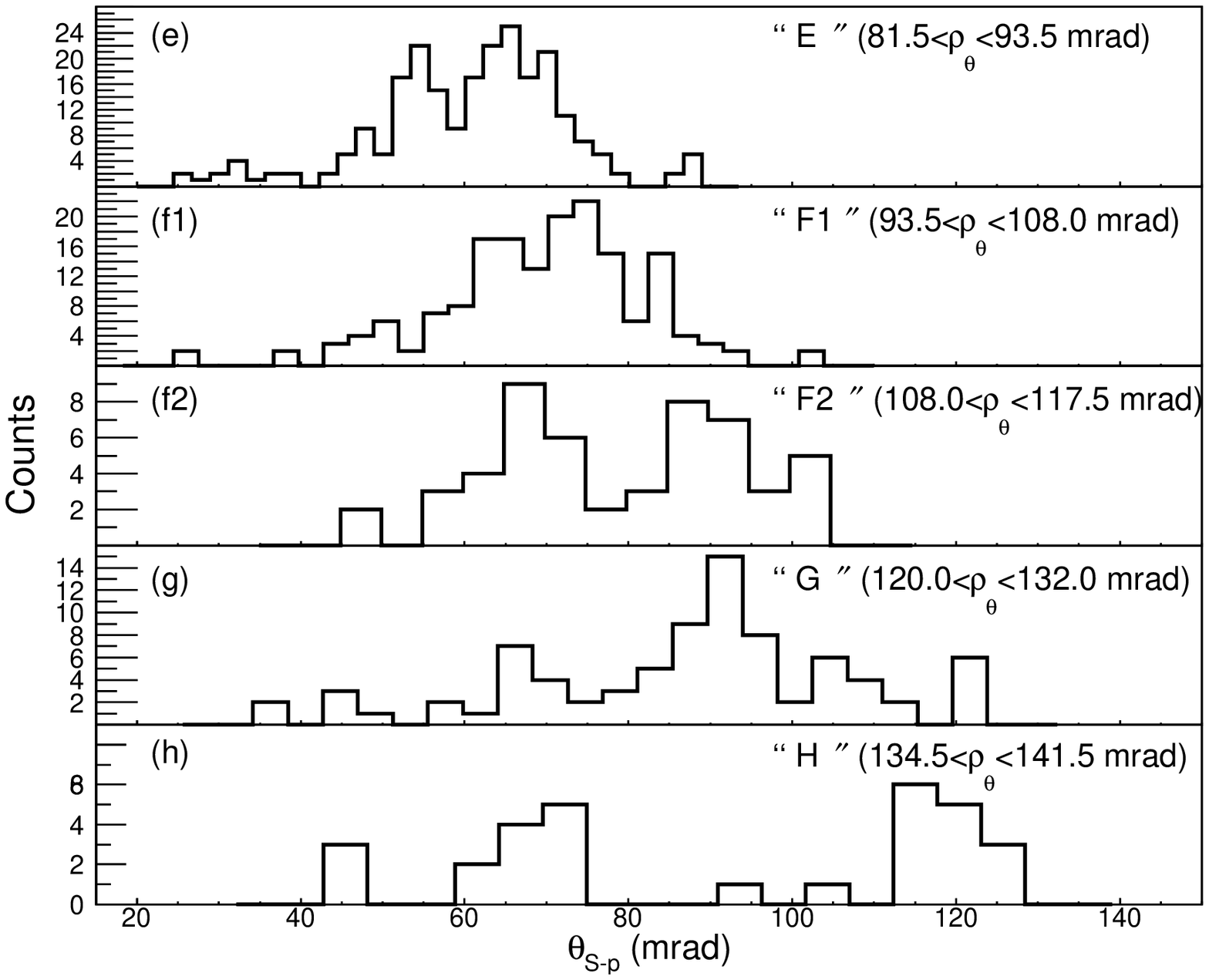}
\end{minipage}
\caption{Angular correlations $\theta_{\rm{S}\text{-}p}$ derived from
the measured $^{28}\rm{S}+\emph{p}+\emph{p}$ coincidences by using the $\rho_\theta$ gates
shown in Fig.~\ref{Theta_rho_Ar}. The corresponding $\rho_\theta$ ranges for the peaks
``A'', ``B'', ``C'', ``D1'', ``D2'', ``E'', ``F1'', ``F2'', ``G'', and ``H'' are
$38.5<\rho_\theta<48.0$ mrad [panel (a)], $48.0<\rho_\theta<60.0$ mrad [panel (b)],
$62.5<\rho_\theta<67.0$ mrad [panel (c)], $67.0<\rho_\theta<72.0$ mrad [panel (d1)],
$72.0<\rho_\theta<81.5$ mrad [panel (d2)], $81.5<\rho_\theta<93.5$ mrad [panel (e)],
$93.5<\rho_\theta<108.0$ mrad [panel (f1)], $108.0<\rho_\theta<117.5$ mrad [panel (f2)],
$120.0<\rho_\theta<132.0$ mrad [panel (g)], and $134.5<\rho_\theta<141.5$ mrad [panel (h)], respectively.}
\label{Theta_Ar_all}
\end{figure*}

Since the $\theta_{\rm{S}\text{-}p}$ distributions reflect energy spectra of protons emitted from the 2\emph{p} decay of $^{30}$Ar states, they provide insight into the decay mechanisms of the parent states. The pattern of the $\theta_{\rm{S}\text{-}p}$ distribution carries information on the decay branches of the $^{30}$Ar state. Fig.~\ref{Theta_Ar_all} displays the $\theta_{\rm{S}\text{-}p}$ spectra obtained from the measured $^{28}\rm{S}+\emph{p}+\emph{p}$ coincidences which are selected by imposing the $\rho_\theta$ gates ``A'', ``B'', ``C'', ``D1'', ``D2'', ``E'', ``F1'', ``F2'', ``G'', and ``H'' shown in Fig.~\ref{Theta_rho_Ar}. The proton spectrum of a simultaneous 2\emph{p} decay of a state exhibits a relatively broad peak which corresponds to the half of the total 2\emph{p}-decay energy $Q_{2p}.$ In the case of the sequential emission of protons, a typical double-peak structure appears in the proton spectrum, and the two peaks are related to the decay energies of two 1\emph{p} decays, i.e., one peak is located at the decay energy of the intermediate state of 1\emph{p}-decay daughter nucleus ($Q_{1p}$) and the other peak is located around the 1\emph{p}-decay energy of the mother nucleus (i.e., $Q_{2p}-Q_{1p}$). Moreover, multiple-peak structures may also be present in the proton spectrum, which indicate two or more decay branches. Therefore, one can obtain hints of the decay mechanism on the basis of the angular $\theta_{\rm{S}\text{-}p}$ distribution. In the case of $^{30}$Ar, one can see in Fig.~\ref{Theta_Ar_all} that except the state ``B'', all other $^{30}$Ar states show two or more $\theta_{\rm{S}\text{-}p}$ peaks, which indicate a sequential decay mechanism. Concerning peak ``B'', the angular $^{28}$S-proton spectrum presents a relatively broad peak which is almost twice more wide than that expected for a simultaneous 2\emph{p} decay. On the other hand, the spectrum does not point to a sequential 2\emph{p} emission, where the typical double-peak structure appears. Such an unexpected pattern was carefully studied in our previous work~\cite{Mukha2015PRL}. There, the peak ``B'' has been assigned to the g.s.~of $^{30}$Ar. Its decay mechanism was identified in a transition region between simultaneous 2\emph{p} decay and sequential emission of protons. The peak ``C'' was regarded as the first excited state of $^{30}$Ar. The decay of this state presents the first hint of a fine structure in the 2\emph{p} decay, which provides the natural interpretation of the peak ``A'' and peak ``C''~\cite{Mukha2015PRL}. In the present work, we will discuss the excited states of $^{30}$Ar with decay energies higher than that of the peak ``C'' and deduce their decay properties and decay mechanisms. Regarding the state ``H'' which is located at about 15 MeV above the 2\emph{p} threshold, we will not discuss it further due to a few decay events observed from this state. Before investigating the 2\emph{p} decay properties of the $^{30}$Ar states, the states of its 1\emph{p} decay daughter nucleus $^{29}$Cl must be studied.

\subsection{Decay energies of low-lying $^{29}$Cl states}\label{subsec:Cl_ex}

\begin{figure}[!htbp]
\centerline{\includegraphics[scale=0.33, angle=0]{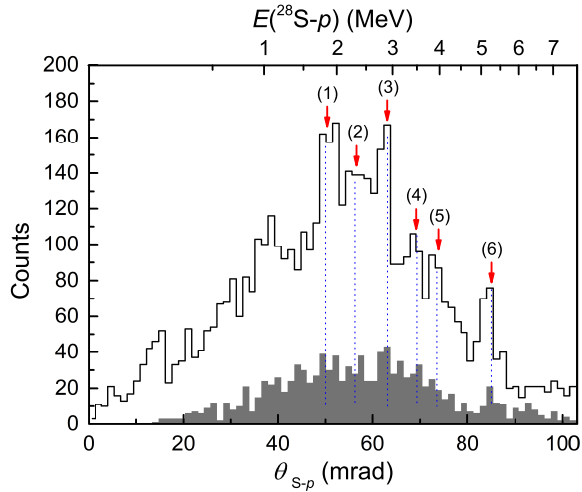}}
\caption{Angular correlations $\theta_{\rm{S}\text{-}p}$ derived from the measured
$^{28}\rm{S}+\emph{p}$ double coincidences (unfilled histogram) and that deduced
from the $^{28}\rm{S}+\emph{p}+\emph{p}$ triple coincidences (grey-filled histogram).
The blue broken lines together with red arrows indicate the peaks which appear
in both histograms, and they suggest the possible $^{29}$Cl resonances,
whose 1\emph{p}-decay energies are shown in the upper axis in MeV.}
\label{Theta_Ar_23}
\end{figure}

As shown in Ref.~\cite{Mukha2015PRL}, the comparison of the $\theta_{\rm{S}\text{-}p}$ distribution obtained from the measured $^{28}\rm{S}+\emph{p}$ double coincidence and that from $^{28}\rm{S}+\emph{p}+\emph{p}$ triple coincidences provide some guidance on the states in $^{29}$Cl, which were populated in the experiment. Such a comparison is displayed in Fig.~\ref{Theta_Ar_23}. In the $^{28}\rm{S}+\emph{p}$ case (the unfilled histogram in Fig.~\ref{Theta_Ar_23}), the $^{29}$Cl states may be populated via several possible reactions on $^{31}$Ar, e.g., the two-step reaction $^{31}\rm{Ar}~\rightarrow~^{30}\rm{Ar}+\emph{n}$ followed by $^{30}\rm{Ar}~\rightarrow~^{29}\rm{Cl}+\emph{p}$, or via the direct fragmentation $^{31}\rm{Ar}~\rightarrow~^{29}\rm{Cl}+\emph{n}+\emph{p}.$ Concerning the $\theta_{\rm{S}\text{-}p}$ spectrum obtained from the $^{28}\rm{S}+\emph{p}+\emph{p}$ coincidence (grey-filled histogram in Fig.~\ref{Theta_Ar_23}), population of $^{29}$Cl states are presumably due to the 2\emph{p} emission from $^{30}$Ar states. Therefore, one may expect that the $\theta_{\rm{S}\text{-}p}$ peaks in both distributions indicate the possible $^{29}$Cl states. In Fig.~\ref{Theta_Ar_23}, several $\theta_{\rm{S}\text{-}p}$ peaks (indicated by arrows) with decent intensities coexist in both spectra, which suggest the possible $^{29}$Cl resonances.

Concerning the 1\emph{p}-decay energies of the $^{29}$Cl states indicated by the arrows in Fig.~\ref{Theta_Ar_23}, one can deduce their values by employing the approximate linear relation between the $\theta_{\rm{S}\text{-}p}$ and $\sqrt{Q_{1p}}$. The energies of the observed $^{29}$Cl levels (1-6) are 1.8(1) MeV, 2.3(1) MeV, $2.9^{+0.2}_{-0.3}$ MeV, $3.5^{+0.4}_{-0.3}$ MeV, $3.9^{+0.6}_{-0.5}$ MeV, and $5.3^{+0.7}_{-0.4}$ MeV, respectively. In order to assign the g.s.~of $^{29}$Cl, the isobaric symmetry of mirror nuclei was considered, see Sec.\ \ref{sec:tes}.

\subsection{Decay schemes of $^{30}$Ar and $^{29}$Cl}\label{subsec:scheme_Ar}

\begin{figure}[!htbp]
\centerline{\includegraphics[scale=0.3, angle=0]{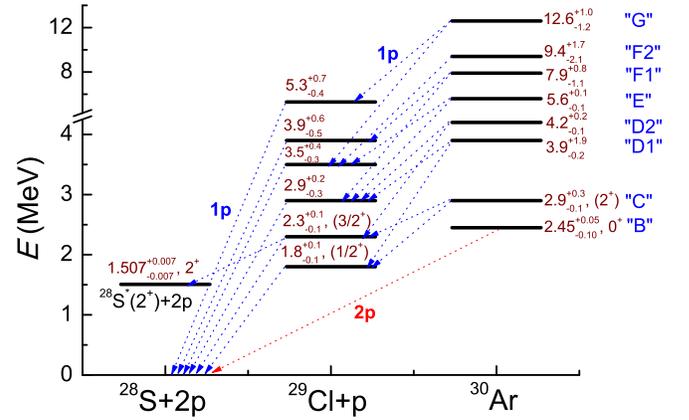}}
\caption{(Color online) Proposed decay schemes of the states observed in $^{30}$Ar and $^{29}$Cl, whose decay energies (in units of MeV) are given relative to their 2\emph{p} and 1\emph{p} thresholds, respectively. The spins and parities given in parentheses are tentative assignments taken from~\cite{Mukha2015PRL}. The energy of $^{28}$S($2^+$) is taken from~\cite{Basunia2013NDS}.}
\label{scheme_Ar}
\end{figure}

As shown in Fig.~\ref{Theta_rho_Ar}, several states of $^{30}$Ar were populated in the present experiment. The decay energy of observed $^{30}$Ar states can be deduced from the $\rho_\theta$ distribution shown in Fig.~\ref{Theta_rho_Ar}(b). The comparison of $\theta_{\rm{S}\text{-}p}$ spectrum obtained from $^{28}\rm{S}+\emph{p}$ coincidences and that deduced from $^{28}\rm{S}+\emph{p}+\emph{p}$ coincidences suggests several states in $^{29}$Cl. By combining these results, we derived a tentative level scheme together with the decay branches for the observed $^{30}$Ar and $^{29}$Cl states. It is shown in Fig.~\ref{scheme_Ar}. The g.s.~and first excited state of $^{30}$Ar and $^{29}$Cl have been discussed in the Ref.~\cite{Mukha2015PRL}. The decay scheme of higher excited states is interpreted below.

\section{Discussion}\label{sec:discuss}

In the previous section a quite detailed energy level and decay scheme was deduced. Such an assignment is based on the limited kinematic information derived from the angular distributions. The data also have limited statistical significance, which is a common situation for extreme exotic nuclear systems near and beyond the driplines. Therefore, the interpretation of the data is partly based on some speculations, which require detailed explanation. Our interpretation of the data is based on reasonable arguments, which take into account present knowledge. It is self-consistent, no alternative self-consistent interpretation which covers all aspects of the observed picture is achieved. Different issues which one had to elaborate in order to arrive to the interpretation shown in Fig.\ \ref{scheme_Ar} are discussed in this section. Some of these issues have already been considered in Ref.\ \cite{Mukha2015PRL}. All arguments are presented below in a systematic way.

\subsection{Signature of $^{30}$Ar ground state}\label{sec:where}

Identification of the $^{29}$Cl and $^{30}$Ar g.s.\ energies is the most important assignment on which the whole interpretation is based. The low-energy peaks in the $^{28}$S-$p$ and $^{28}$S-$p$-$p$ correlations spectra may arise from decay channels populating the excited states of $^{28}\mbox{S}^*$, which are instantaneously de-excited by $\gamma$-emission. In the present experiment the reaction target area was observed by an ancillary $\gamma$-ray detector. With its total registration efficiency of about $5\%$, this information could be useful for counting rates, say, an order of the magnitude higher than those available for $^{29}$Cl and $^{30}$Ar.

The candidates for $^{30}$Ar g.s.\ in Fig.\ \ref{Theta_rho_Ar} are peak ``A'' at $\rho_{\theta} \sim 44$ mrad (corresponding to $Q_{2p}=1.4$ MeV) and peak ``B'' at $\rho_{\theta} \sim 59$ mrad (corresponding to $Q_{2p}=2.45$ MeV). We point to three reasons which make the peak ``B'' preferable prescription for the $^{30}$Ar ground state.

The first argument is connected with a population cross section for the peak ``A''. It comprises less than $5\%$ of the total population intensity of all $^{30}$Ar states, and this value is unexpectedly low for the ground state. For comparison, one may look at some examples of corresponding values obtained in the nucleon-knockout experiments populating $s$-$d$ shell nuclei beyond the dripline. They are: $\sim20\%$ for $^{19}$Mg\ \cite{Mukha2007PRL}, $\sim35\%$ for $^{16}$Ne\ \cite{Brown2015PRC}, $\sim60\%$ for $^{26}$O\ \cite{Kondo2016PRL}. These values and also other examples, which can be found in the literature, vary quite broadly demonstrating strong sensitivity to the individual structure of the precursor nucleus. However, such values never seem to be extremely small.

The second argument is connected with systematics of odd-even energy staggering (OES) and proton correlation pattern for the peak ``A''. The OES is defined as
\[
2E_{\mathrm{OES}} = S^{(A)}_{2N}-2S^{(A-1)}_{N}\,,
\]
where $S^{(A)}_{2N}$ and $2S^{(A-1)}_{N}$ are separation energies for two nucleons in the system with mass number $A$ and for one nucleon in its $A-1$ subsystem. This can be interpreted as a phenomenological paring energy value computed with the assumption that the structure of the nuclear with mass number $A$ is represented by two valence nucleons populating single-particle configurations corresponding to $A-1$ system ground state. It was demonstrated in Fig.\ 6 of Ref.\ \cite{Mukha2015PRL} that the systematics of OES is very similar for the isotone chain leading to $^{30}$Ar and its mirror $^{30}$Mg isobar. The extrapolated value for $^{30}$Ar is $2E_{\mathrm{OES}} = 2.25$ MeV. However, it is known that such a systematics breaks near the borderline of the nucleon stability\ \cite{Comay1988PLB}. Therefore, the actually expected value of $2E_{\mathrm{OES}}$ should be a factor $0.4-0.7$ smaller.

Several reasonable prescriptions for proton and two-proton decay energies of $^{29}$Cl and $^{30}$Ar, respectively, are shown in Table \ref{tab:oes}. By considering the $^{28}$S-$p$ angular correlations in Fig.\ \ref{Theta_Ar_all}, the double-peak structure for the 1.4 MeV peak ``A'' can be interpreted as the result of the sequential emission of two protons with the energies of 0.6 and 0.8 MeV. As far as the emission order of protons is not known, one has to consider both prescriptions for the $^{29}$Cl $Q_p$ value, which is marked as P1 and P2 in Table \ref{tab:oes}. Both variants provide the OES values far beyond the range expected from systematics. Let's assume that the double-peak structure of the $p$-$p$ correlations is actually connected with a statistical ``staggering'' due to low data statistics and in reality a single peak characteristic for ``true'' $2p$ emission should exist. Then one should assume the higher reasonable $^{29}$Cl g.s.\ position which value can be found as $Q_p=1.8$ MeV from Fig.\ \ref{Theta_Ar_23}. For this prescription, marked as P3, the OES value is somewhat overestimated. If one correlates the $^{30}$Ar g.s.\ with the peak ``B'' in Fig.\ \ref{Theta_rho_Ar} (at $Q_{2p}=2.45$ MeV), and assume $Q_p=1.8$ MeV (the prescription P4), then the obtained OES value nicely fits the expected range. The relative population intensity of the peak ``B'' ($\sim 15 \%$ of the total) is also reasonably consistent with expectation for the ground state. An additional argument for the choice of $Q_p=1.8$ MeV prescription comes from systematics of Coulomb displacement energies is described in the next section.

\begin{table}[b]
\caption{The odd-even staggering energy in $^{30}$Ar calculated under different assumptions about proton and two-proton decay energies in $^{29}$Cl and $^{30}$Ar.}
\begin{ruledtabular}
\begin{tabular}[c]{cccccc}
  & Expected &  P1 & P2 & P3 & P4  \\
\hline
$Q_{2p}$($^{30}$Ar) &          & 1.4  & 1.4 & 1.4 & 2.45  \\
$Q_{p}$($^{29}$Cl)  &          & 0.6  & 0.8 & 1.8 &  1.8  \\
$2E_{\mathrm{OES}}$ & 0.9--1.6 & -0.2 & 0.2 & 2.2 &  1.15  \\
\end{tabular}
\end{ruledtabular}
\label{tab:oes}
\end{table}

The plausible explanation of the  ``A'' structure at 1.4 MeV is that this is a ``fine-structure peak'' for $2p$ emission from the first excited (probably $2^+$) state in $^{30}$Ar into the first excited state of $^{28}$S ($2^+$ at 1.507 MeV, see Fig.\ \ref{scheme_Ar}). The further discussion of this issue needed for a consistent description of the $^{29}$Cl and $^{30}$Ar decay schemes (see, Fig.\ \ref{scheme_Ar}) is provided in the Sec.\ \ref{sec:spec_struct}.

\subsection{Thomas-Ehrman effect in $^{29}$Cl-$^{29}$Mg}\label{sec:tes}

\begin{figure}[!htbp]
\centerline{\includegraphics[width=0.48\textwidth]{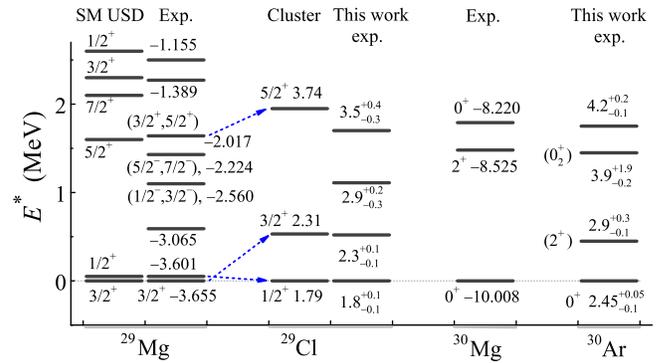}}
\caption{Level schemes of $^{29}$Cl and $^{30}$Ar compared with schemes of isobaric mirror partners. Shell-model predictions (labeled with SM USD) for $^{29}$Mg and results of a cluster potential model (labeled with Cluster) for $^{29}$Cl are compared with the experimental data. The energies of $^{29}$Cl are shown relative to the 1\emph{p} threshold. The energies of $^{29}$Mg states were shifted down by 3.655 MeV in order to compare them with the mirror states in $^{29}$Cl.}
\label{Model_mirror}
\end{figure}

If one assume the single-particle nature of the $^{29}$Cl low-lying states, their energies can be reliably evaluated basing on the spectrum of the isobaric mirror partner, the $^{29}$Mg nucleus. The g.s.\ of $^{29}$Mg is known to have spin-parity $3/2^+$ \cite{NNDC} and it is reasonable to assume the single-particle $d$-wave structure of this state. The first excited state is separated just by 54 keV, see Fig.\ \ref{Model_mirror}. According to shell model calculations with Brown-Wildenthal USDB Hamiltonian \cite{Brown2006PRC}, this can be expected to be an $s$-wave $1/2^+$ state. For such a situation of practically degenerated $s$- and $d$-wave states in $^{29}$Mg, one may expect a strong modification of the level scheme due to the Thomas-Ehrman effect \cite{Thomas1952PR,Ehrman1951PR}, which can be evaluated by using a simple potential model.

There are two major parameters, which controls the Coulomb displacement energies of nucleons in the potential model: (i) the charge radius of the core nucleus, which is a major characteristic of the charge distribution, and (ii) the radius parameter of the potential, which controls the nucleon orbital size. For the $^{28}$S-$p$ channel, the Woods-Saxon potential is used with two parameter sets, see Table \ref{tab:potentials}. The first set (P1 and P2 cases) is quite typical for light nuclei, it was also employed in the work on $2p$ radioactivity \cite{Pfutzner2012RMP}. The second set (P3--P5) uses potential parameters of a global parameterization destined to obtain single-particle states for shell-model calculations \cite{Schwierz2007arXiv}. The charge radius of $^{28}$S is not known but can be extrapolated using the known values for $^{32-36}$S \cite{Angeli2013ADNDT}, which are in the range 3.26--3.30 fm. We consider the range $r_{\text{ch}} = 3.18 - 3.26$ fm as a realistic value for $^{28}$S. For each nuclear potential set we use the upper and lower charge radius values (see Table \ref{tab:potentials}) to define the Coulomb potential of the homogeneously charged sphere of radius $r_{\text{sp}}$ by the following expression
\[
r^2_{\text{sp}} = (5/3) \left[r^2_{\text{ch}}  + r^2_{\text{ch}}(p)\right]  \,.
\]
where $r_{\text{ch}}(p)=0.8$ fm is the proton charge radius. The obtained $Q_p$ values are in the range $1.69 - 1.79$ MeV, which is consistent with the prescription of $Q_p=1.8$ MeV for the peak denoted ``(1)'' in Fig.\ \ref{Theta_Ar_23} in order to correspond to $^{29}$Cl ground state.

One may evaluate how strongly should one modify the potential model input to obtain $Q_p=0.8$ MeV or $Q_p=0.6$ which is required to associate the 1.4 MeV peak ``A'' with the $^{30}$Ar g.s. The P5 case in Table \ref{tab:potentials} shows how large should be the charge radius of $^{28}$S in order to get the $Q_p$ value within the above-mentioned range. It is found that a value as large as $r_{\text{ch}}> 5$ fm is needed. However, charge radii as large as 5 fm become available for nuclei heavier than neodymium and mass numbers twice larger than that of $^{30}$Ar. Thus the decay energies $Q_p \sim 0.6-0.8 $ MeV are unrealistic for $^{29}$Cl.

\begin{table}[b]
\caption{Parameters of the $^{28}$S-$p$ potentials for $1/2^+$ state. On the neutron-rich side of the isobar these potentials exactly reproduce the neutron separation energy $S_n=3.601$ MeV for the $^{29}$Mg first excited state.}
\begin{ruledtabular}
\begin{tabular}[c]{cccccc}
  & P1 & P2  & P3  & P4 & P5 \\
\hline
$r_0$, $a$, fm  & \multicolumn{2}{c}{1.2, 0.65} & \multicolumn{3}{c}{1.26, 0.662} \\
$r_{\text{ch}}$, fm  & 3.18 & 3.26 & 3.18 & 3.26 & 5.0 \\
$U_0$, MeV & \multicolumn{2}{c}{$-41.866$} & \multicolumn{3}{c}{$-38.836$} \\
$Q_p$, MeV & 1.785 & 1.755 & 1.715 & 1.685 & 0.818 \\
\end{tabular}
\end{ruledtabular}
\label{tab:potentials}
\end{table}

The $^{29}$Mg ground $3/2^+$ state and the 1.638 keV state (which may be assumed to be $5/2^+$ according to shell model calculations) belong to a $d$-wave doublet (it will be shown below that the actual situation may be more complicated, see discussion in Sec.\ \ref{sec:spec_struct}). The states with such a structure have considerably larger ($\sim 0.5$ MeV) Coulomb displacement energies in comparison with the $s$-wave $1/2^+$ state, see Fig.\ \ref{scheme_Ar}, and thus they provide different level ordering in $^{29}$Cl compared to $^{29}$Mg. Then the $3/2^+$ prescription for peak ``(2)'' with $Q_p=2.3$ MeV, and the $5/2^+$ prescription for peak ``(4)'' with $Q_p=3.5$ MeV are possible as well (see Fig.\ \ref{Theta_Ar_23}).

Fig.\ \ref{scheme_Ar} shows the level scheme of $^{30}$Ar compared to that of the isobaric mirror partner $^{30}$Mg. There is an important difference in these schemes, which could be an evidence for strong Thomas-Ehrman effect in this isobaric mirror pair as well. Another origin could be a quite specific structure of the first excited state in $^{30}$Ar, as it is argued in Sec.\ \ref{sec:spec_struct}.

\subsection{Transition dynamics of $^{30}$Ar ground state decay}\label{subsec:transition}

As we have shown above, the assignment of peak ``B'' in Fig.\ \ref{Theta_rho_Ar} to $^{30}$Ar ground state is plausible from the point of view of different energy systematics. However, the corresponding $^{28}$S-$p$ angular correlations of this peak [see Fig.\ \ref{Theta_Ar_all} (b)] show the pattern which we first found problematic to interpret. There is neither a single narrow ``central'' peak, typical for ``true'' $2p$ emission, nor a double (or other even-number) peak structure associated with the sequential emission of protons. Estimates of Ref.\ \cite{Mukha2015PRL} demonstrated that the natural explanation of this fact is connected with the peculiar ``transitional'' decay dynamics which is exactly on the borderline between ``true'' $2p$ and sequential $2p$ decay. The behavior of physical observables in such a ``transitional'' region demonstrates features analogous to phase transitions. Namely there is a very high sensitivity of observables to minor variations of parameters. The parameters of nuclear decays are not subject of free variation from outside like in phase transitions considered e.g.\ in thermodynamics. However, if our system of interest appears to belong to ``transitional'' decay dynamics case, then strong sensitivity to parameters paves the way to precise determination of parameters (or some of their combinations) based on observables.

\begin{figure}[tb]
\centerline{\includegraphics[width=0.48\textwidth]{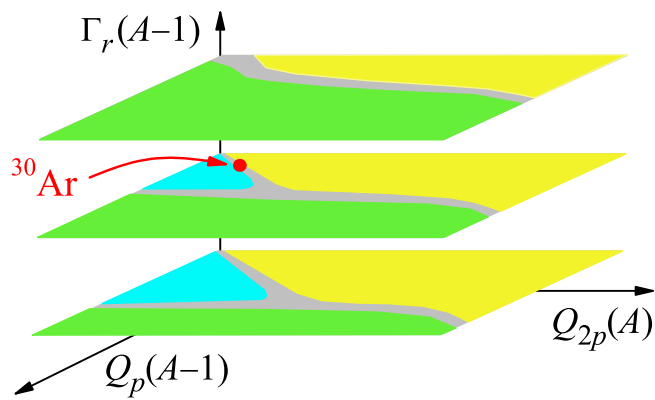}}
\caption{(Color online) Areas of dominance of different decay mechanisms of $2p$ precursors with mass number $A$ dependance of general decay parameters $Q_{2p}{(A)}$, $Q_p{(A-1)}$, and $\Gamma_r{(A-1)}$. The blue-, green- and yellow-filled areas correspond to the dominating true, democratic and sequential $2p$-decay mechanisms, respectively. The transition regions are indicated by the grey areas. The $^{30}$Ar decay (red dot) is located in the transition region.}
\label{transit}
\end{figure}

Transitions between different regimes of three-body decays have been discussed in detail in Ref.\ \cite{Golubkova2016PLB}. There exist three distinct mechanisms of such decays --- ``true'', democratic, and sequential --- which all are characterized by the distinct pictures of three-body correlations and different systematic of the lifetimes Ref.\ \cite{Pfutzner2012RMP}. In the most common case the transitions between these regimes are defined by three parameters: three-body decay energy $Q_{2p}{(A)}$, two-body decay energy of the ground state in the core-$p$ subsystem $Q_p{(A-1)}$, and the width of the core-$p$ ground state resonance $\Gamma_r{(A-1)}$. Qualitative illustration of the transition phenomenon in the three-body systems is provided in Fig.\ \ref{transit}.

\begin{figure}[tb]
\centerline{\includegraphics[width=0.43\textwidth]{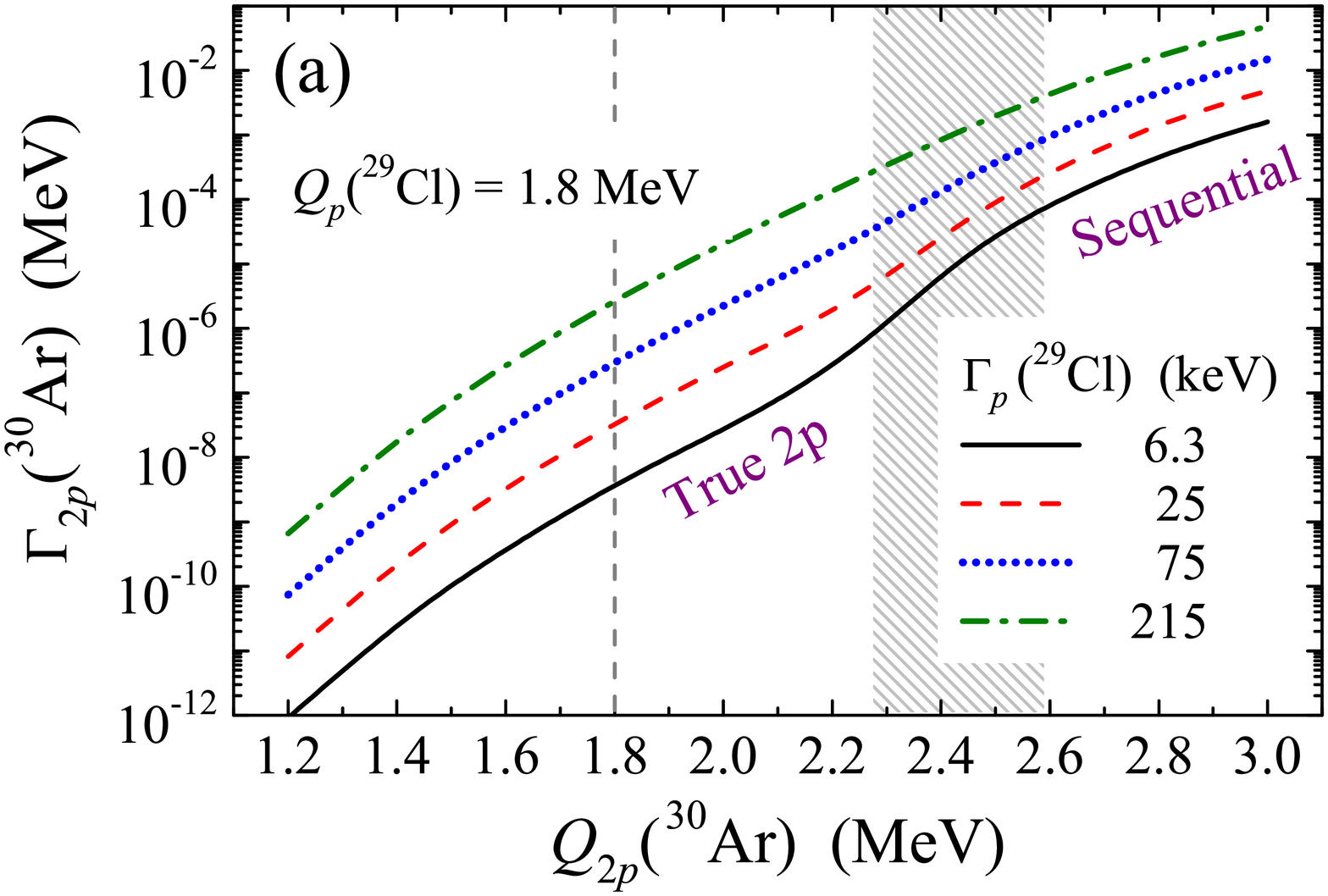}}
\centerline{\includegraphics[width=0.43\textwidth]{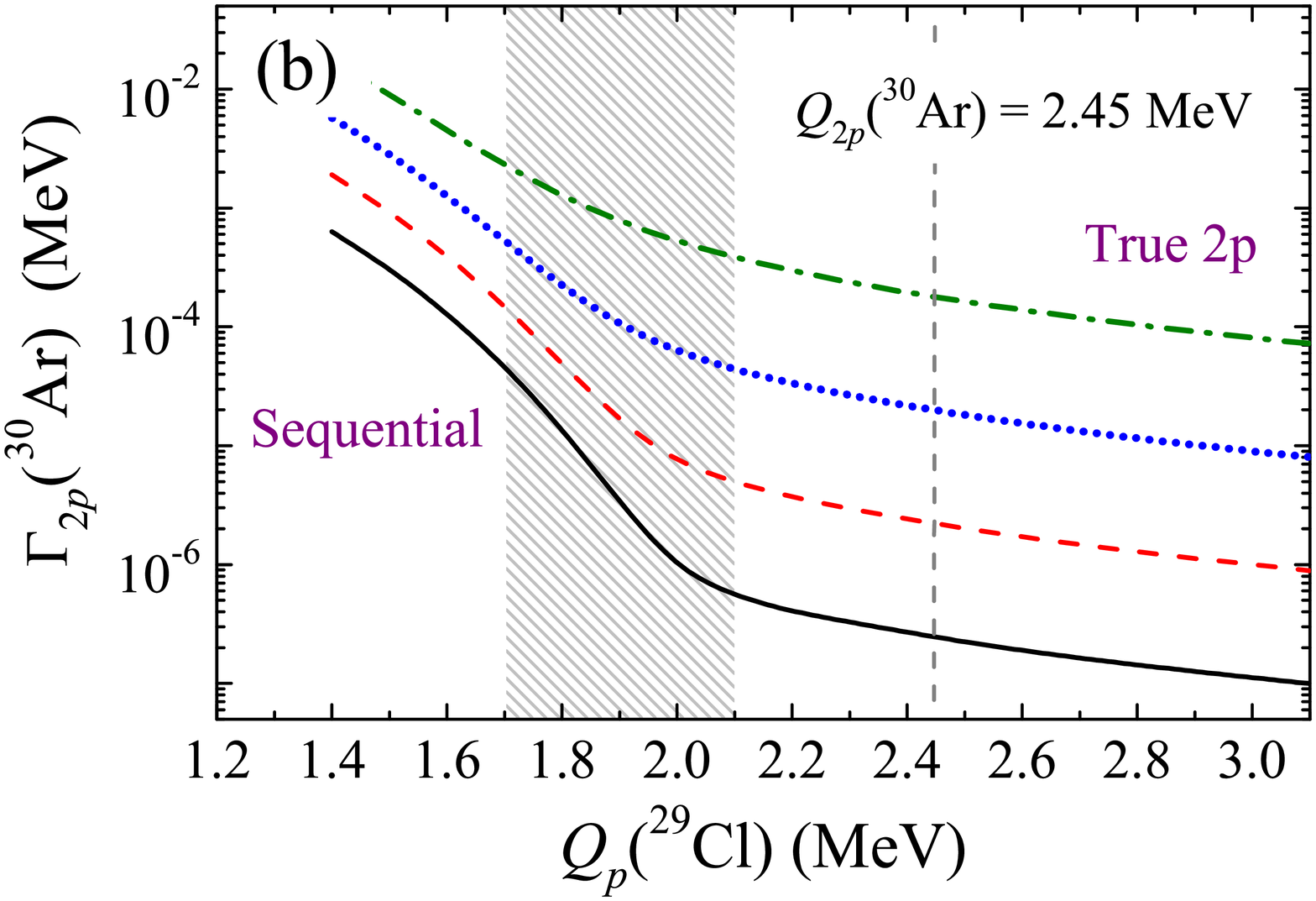}}
\caption{(Color online) Two-proton decay width $\Gamma_{2p}$ of $^{30}$Ar around the transition regions between the true $2p$ and sequential $2p$ emission mechanisms. Panel (a) shows the width as function of the 2p-decay energy $Q_{2p}$($^{30}$Ar) for a fixed value of $Q_p(^{29}\text{Cl}) = 1.8$ MeV. Panel (b) shows the width of $^{30}$Ar g.s.\ as function of $1p$-decay energy of the $^{29}$Cl subsystem, $Q_p$($^{29}$Cl) for the fixed value of $Q_{2p}(^{30}\text{Ar}) = 2.45$ MeV. The solid, dashed, dotted, and dash-dotted curves correspond to the $\Gamma_p(^{29}\text{Cl})$ values of 6.3, 25, 75, and 225 keV, respectively, at $Q_p(^{29}\text{Cl}) = 1.8$ MeV. The hatched areas indicate the transition regions.}
\label{transit-width}
\end{figure}

Based on the direct-decay model which was improved in the content of the present work, theoretical and simulation studies of the $^{30}$Ar $2p$ decay dependence on three general decay parameters have been performed in \cite{Golubkova2016PLB}. The strong dependence of lifetime systematics in the transition region on the parameter values is illustrated in Fig.\ \ref{transit-width}. The lifetime curves demonstrate a kink at the transition situation. This kink is more expressed for small values of the two-body widths. Different curves are evaluated based on different assumption about proton width of the $^{29}$Cl g.s. The gap between lifetime curves is much larger in the true $2p$ decay part of the plots. This is connected with the fact that the sequential $2p$ decay width linearly depends on the $1p$ $^{29}$Cl g.s.\ width, while the true $2p$ decay width depend on that quadratically. Consequently, a stronger kink for smaller $1p$ width values is needed to compensate for this effect in the transition region.

\begin{figure}[!htbp]
\centerline{\includegraphics[scale=0.40, angle=0]{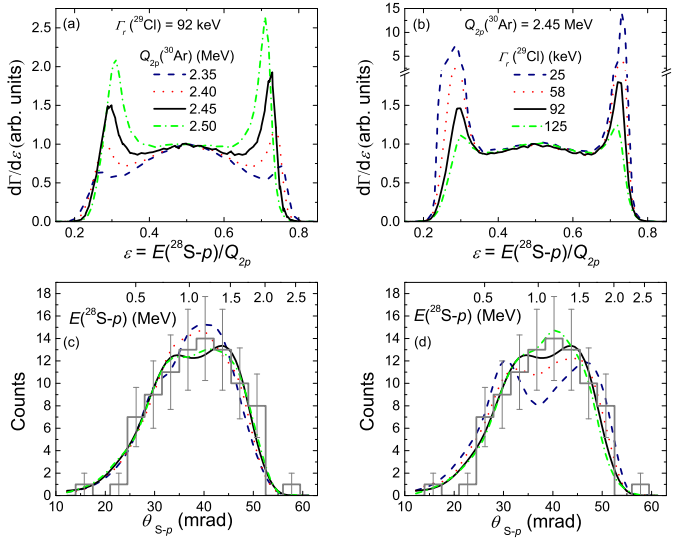}}
\caption{(Color online) Transition from the true-2\emph{p} decay mechanism to the sequential proton emission mechanism
of $^{30}$Ar ground state. (a) Energy distribution between core and one of the protons calculated by employing
improved direct decay model \cite{Golubkova2016PLB}, where the 2\emph{p}-decay energy of $^{30}$Ar is varied.
(b) Same as (a) but with variation of the width of $^{29}$Cl g.s.~In the panels (c) and (d), the
experimental $\theta_{\rm{S}\text{-}p}$ distribution measured for the decay of $^{30}$Ar g.s.~(grey histogram
with statistical uncertainties) is compared with those stemming from respective theoretical distributions in
panels (a) and (b)after experimental bias is taken into account via Monte Carlo simulations.}
\label{Ar30_gs_dependence}
\end{figure}

The observables, which were found of considerable practical interest in the context of the present work, are three-body correlations among the heavy fragment and the protons. Fig.~\ref{Ar30_gs_dependence} (a) displays the calculated energy distributions between the $^{28}$S and one of the emitted protons. In the calculations, the resonant energy of $^{29}$Cl g.s.,~$Q_{p}(^{29}\rm{Cl})$ is set to 1.8 MeV and the $\Gamma(^{29}\rm{Cl})$ is fixed at 92 keV. One may clearly see that the shape and width of the spectrum profile change dramatically with the variation of $Q_{2p}(^{30}\rm{Ar})$, which represents a strong sensitivity of the decay mechanism to the $Q_{2p}(^{30}\rm{Ar})$. In the case of small $Q_{2p}(^{30}\rm{Ar})$, e.g., 2.35 MeV, the energy distribution between $^{28}$S and proton [blue dashed curve in Fig.~\ref{Ar30_gs_dependence} (a)] is mainly characterized by a bell-like spectrum centered at $\varepsilon=0.5$, which indicates the true 2\emph{p} decay. In contrast, the spectrum with a bit larger $Q_{2p}(^{30}\rm{Ar})$ value [e.g., $Q_{2p}(^{30}\rm{Ar})=2.50$ MeV, the green dashed-dotted curve in Fig.~\ref{Ar30_gs_dependence}(a)] is mainly featured by a double-peak pattern (with two peaks at $\varepsilon=0.75$ and at $\varepsilon=0.25$), which typically corresponds to the sequential 2\emph{p} emission. Therefore, the correlation pattern is extremely sensitive to calculation parameters, where small variations of $Q_{2p}(^{30}\rm{Ar})$ cause dramatic changes of the shapes of distributions. Similarly, the sensitivity of the energy distribution to $\Gamma(^{29}\rm{Cl})$ was also investigated. The corresponding results are shown in Fig.~\ref{Ar30_gs_dependence}(b). Here $Q_{2p}(^{30}\rm{Ar})=$ 2.45 MeV and $Q_{p}(^{29}\rm{Cl})=$ 1.8 MeV. With the increase of $\Gamma(^{29}\rm{Cl})$, an obvious change from a sequential two-body decay case to a true three-body decay situation can be observed.

In order to compare the model predictions of the $^{28}$S-\emph{p} angular correlations with the experimental data, Monte Carlo simulations of the detector response to the 2\emph{p} decay of $^{30}$Ar g.s.~were performed. The momenta of three decay products used in the simulations were taken from the predictions of the direct decay model~\cite{Golubkova2016PLB}. The corresponding results are shown in panels (c) and (d) of Fig.~\ref{Ar30_gs_dependence}, which illustrate the dependence of the simulated $\theta_{\rm{S}\text{-}p}$ spectrum on $Q_{2p}(^{30}\rm{Ar})$ and $\Gamma(^{29}\rm{Cl})$, respectively. In comparison of experimental $\theta_{\rm{S}\text{-}p}$ distribution (grey histogram with statistical uncertainties) and various simulations, the model prediction with $Q_{2p}=2.45$ MeV and $\Gamma(^{29}\rm{Cl})=92$ keV (black solid curves in panels (c) and (d) of Fig.~\ref{Ar30_gs_dependence}) reproduces the data.

The sensitivity of energy distributions of decay products to reasonable combinations of the parameters \{$Q_{2p}(^{30}\rm{Ar})$, $Q_{p}(^{29}\rm{Cl})$, $\Gamma_r(^{29}\rm{Cl})$\} was systematically investigated in \cite{Golubkova2016PLB}. A statistical analysis allowed to find the preferable combination of three parameters: $Q_{2p}(^{30}\rm{Ar}) = 2.45^{+0.05}_{-0.10}$ MeV and $Q_{p}(^{29}\rm{Cl}) = 1.8 \pm 0.1$ MeV, $\Gamma_{r}(^{29}\rm{Cl}) = 85\pm 30$ keV. The result $Q_{2p}(^{30}\rm{Ar}) = 2.45^{+0.05}_{-0.10}$ MeV is somewhat different compared to the first-reported value $Q_{2p}(^{30}\rm{Ar}) = 2.25^{+0.15}_{-0.10}$ MeV from Ref.\ \cite{Mukha2015PRL}, but consistent within the error bars. The determination of the width $\Gamma(^{29}\rm{Cl})$ in an indirect way, based on the $^{30}$Ar correlation data, should be regarded as a novel result of the proposed approach.

Three aspects of the current analysis should be emphasized. (i) The current studies were performed based on the data with quite limited statistics. Just fact of such opportunity is already very encouraging. Thus they should be regarded more as a proof-of-concept rather than a final result. It was not evident in advance that differences in the observed patterns would be sufficient to put restrictions on the physical parameters. (ii) It is demonstrated that the method is working even utilizing kinematically very limited information (angular distributions in the (heavy ion)-$p$ channel. The application of the method to complete kinematics where the momenta vectors of all outgoing decay products are measured, should produce results of higher precision. (iii) To perform width measurements, standard experiments with quite high statistics are required. For example, the determination of $\Gamma(^{29}\rm{Cl})$ in a standard RIB experiment on resonance scattering of $^{28}$S on a hydrogen target would require the availability of a quite intense $^{28}$S beam and registration of hundreds or thousands of decay events. The sensitivity of the proposed method even for low statistics can be understood as a result of kind of ``quantum amplification'': the observed spectrum is not $^{29}$Cl spectrum by itself, but the result of interference of $^{29}$Cl decay amplitudes with other amplitudes involved in $^{30}$Ar decay.

\subsection{Structure of the first excited states in $^{29}$Cl and $^{30}$Ar}
\label{sec:spec_struct}

It was mentioned in Sec.\ \ref{sec:where} that it could be reasonable to assume that the 1.4 MeV ``A'' structure in Fig.\ \ref{Theta_Ar_all} is connected with $2p$ emission from the first excited (probably $2^+$) $^{30}$Ar state into the first excited state of $^{28}$S ($2^+$ at 1.507 MeV, see Fig.\ \ref{scheme_Ar}). We demonstrate in this section that such an assumption leads to strong restrictions on the structure of the first excited states both in $^{29}$Cl and $^{30}$Ar as well as to a consistent description of the decay scheme for the low-lying states in $^{29}$Cl and $^{30}$Ar.

Within the above assumption one may associate the whole feeding to the peak ``A'' with the decay sequence
\begin{equation}
^{30}\text{Ar(2.9)} \rightarrow \, ^{29}\text{Cl(2.3)} \, \rightarrow \,  ^{28}\text{S(1.5)}\, ,
\label{eq:s1}
\end{equation}
while about 3/4 of the peak ``C'' corresponds to the sequence
\begin{equation}
^{30}\text{Ar(2.9)} \, \rightarrow \, ^{29}\text{Cl(2.3)} \, \rightarrow \, ^{28}\text{S(0)} \,.
\label{eq:s2}
\end{equation}
If one compares population intensities in Fig.\ \ref{Theta_Ar_all} (a) and (c), then the sequence (\ref{eq:s2}) is about 2 times more intense than sequence (\ref{eq:s1}). If one sticks to the $3/2^+$ prescription for the 2.3 MeV $^{29}$Cl state then the structure of its wave function (WF) can be schematically presented as
\begin{eqnarray}
\Psi_{^{29}\text{Cl}}(3/2^+) & = & \tilde{\alpha} \Psi_{^{28}\text{S}} (0^+)
[d]_{3/2} + \tilde{\beta} \Psi_{^{28}\text{S}}(2^+)[s]_{1/2}  \nonumber \\
& + & \tilde{\gamma} \Psi_{^{28}\text{S}}(2^+)[d]_{3/2} \, .
\label{eq:struc-29cl}
\end{eqnarray}
The estimate of the penetration factors $P_l(E)$ provides the ratio
\begin{eqnarray}
\frac{P_d(2.3)}{P_s(0.9)} \approx 15 - 25 \,. \nonumber
\label{eq:P_ratio}
\end{eqnarray}
This means that the $\tilde{\beta}$ coefficient should be as large as $0.86 - 0.93$. Therefore, the structure of the first excited state of $^{29}$Cl is totally dominated by the WF component with the $^{28}$S subsystem in the $2^+$ state (the availability of some $\tilde{\gamma}$ terms can only increase the above estimate).

Next we must check whether this assumption is consistent with the decay scheme of 2.9 MeV state of $^{30}$Ar. For the decay of the $E_T=2.9$ MeV state there are two branches: (i) via the 1.8 MeV g.s.\ of $^{29}$Cl (assumed to be $1/2^+$) and (ii) via the 2.3 MeV state of $^{29}$Cl (assumed to be $3/2^+$). These branches are populated with a ratio around 3:1, see Fig.\ \ref{Theta_Ar_all} (c). Let's assume that $E_T=2.9$ MeV is a $2^+$ state. In a schematic notation the $2^+$ state WF can be represented as
\begin{eqnarray}
\Psi_{^{30}\text{Ar}}(2^+)  =  \alpha \Psi_{^{28}\text{S}} (0^+) [sd]_2
+ \beta \Psi_{^{28}\text{S}}(0^+)[d^2]_2 \nonumber \\
 + \gamma \Psi_{^{28}\text{S}}(2^+)[s^2]_0 + \delta \Psi_{^{28}\text{S}}(2^+)[sd]_2 + \epsilon \Psi_{^{28}\text{S}}(2^+)[d^2]_2 \,. \nonumber
\end{eqnarray}
The estimate of penetration factors for the most probable decay branches provides
\[
\frac{P_d(1.1)}{P_s(0.6)} \approx 5 - 8 \,.
\]
To match the observed $d/s$ ratio around 3, one should assume
\[
\frac{\gamma^2+\delta^2}{\alpha^2+\beta^2} \approx 3,
\]
which actually means that the weight of the excited $^{28}$S configuration in the structure of the 2.9 MeV state of $^{30}$Ar is larger than 0.75 (the decay schemes cannot provide information about $\epsilon$ coefficient). The large weight of the excited $^{28}$S configuration in the structure of this state is also required to explain transitions to the 2.3 MeV state of $^{29}$Cl, which is dominated by the excited $^{28}$S configuration as well.

The structure of the $^{29}$Cl first excited state discussed here seems to contradict the discussion of Sec.\ \ref{sec:tes} about Thomas-Ehrman shifts: the valence nucleon can be in a $s$-wave configuration relative to the core (coefficient $\tilde{\beta}$) in Eq.\ (\ref{eq:struc-29cl}). This should produce much smaller TES values. There may be two explanations here. One is that the term with $\tilde{\gamma}$ is dominant in the structure of the $^{29}$Cl $3/2^+$. The other point is that the s-wave component should be much more compact in the WF (\ref{eq:struc-29cl}), because the state is more bound (due to the excitation energy of $^{28}$S $2^+$).

The decay of the $E_T=3.4$ MeV state via the 1.8 MeV g.s.\ of $^{29}$Cl is dominant, while the decay path via the 2.3 MeV state is suppressed. This situation is naturally explained by the assumption that the $0^+_2$ state decays via the $1/2^+$ g.s.\ of $^{29}$Cl by the emission of a $s$-wave proton. The assumption of $0^+$, $2^+$, and $0^+_2$ level ordering in $^{30}$Ar is in agreement with the ordering expected from isobaric symmetry based on the $^{30}$Mg level scheme, see Fig.\ \ref{Model_mirror}. All other possible prescriptions of spins and energies of the  $E_T=2.9$ and $E_T=3.4$ MeV state fail to describe the overall situation with reasonable physics assumptions.

\subsection{Sequential emission of protons from higher excited states in $^{30}$Ar}
\label{sec:seq_ar_high}

\begin{figure}[!htbp]
\centering
\begin{minipage}[b]{0.95\linewidth}
\centering
\includegraphics[scale=0.38, angle=0]{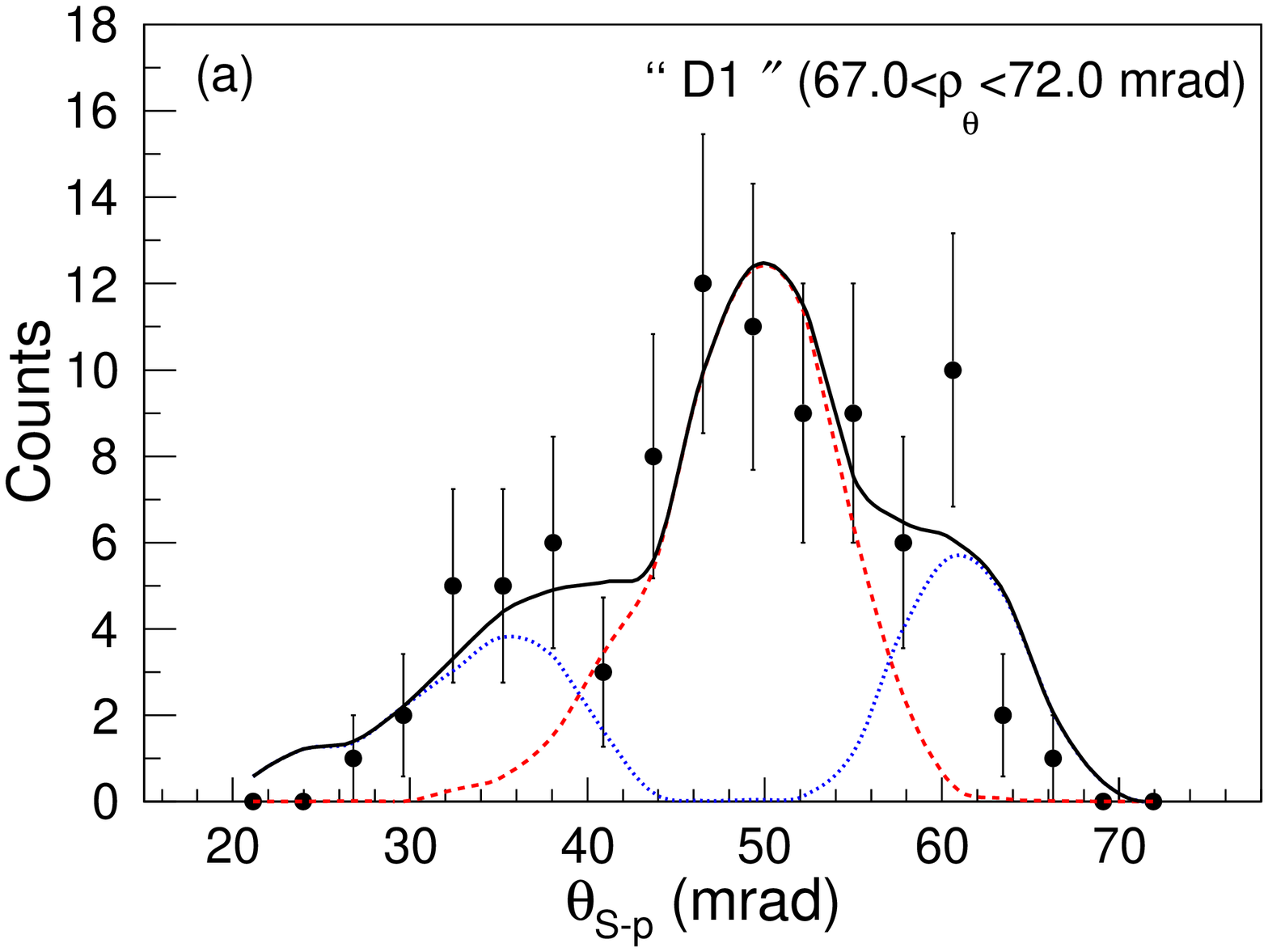}
\end{minipage}
\hspace{1em}
\begin{minipage}[b]{0.95\linewidth}
\centering
\includegraphics[scale=0.38, angle=0]{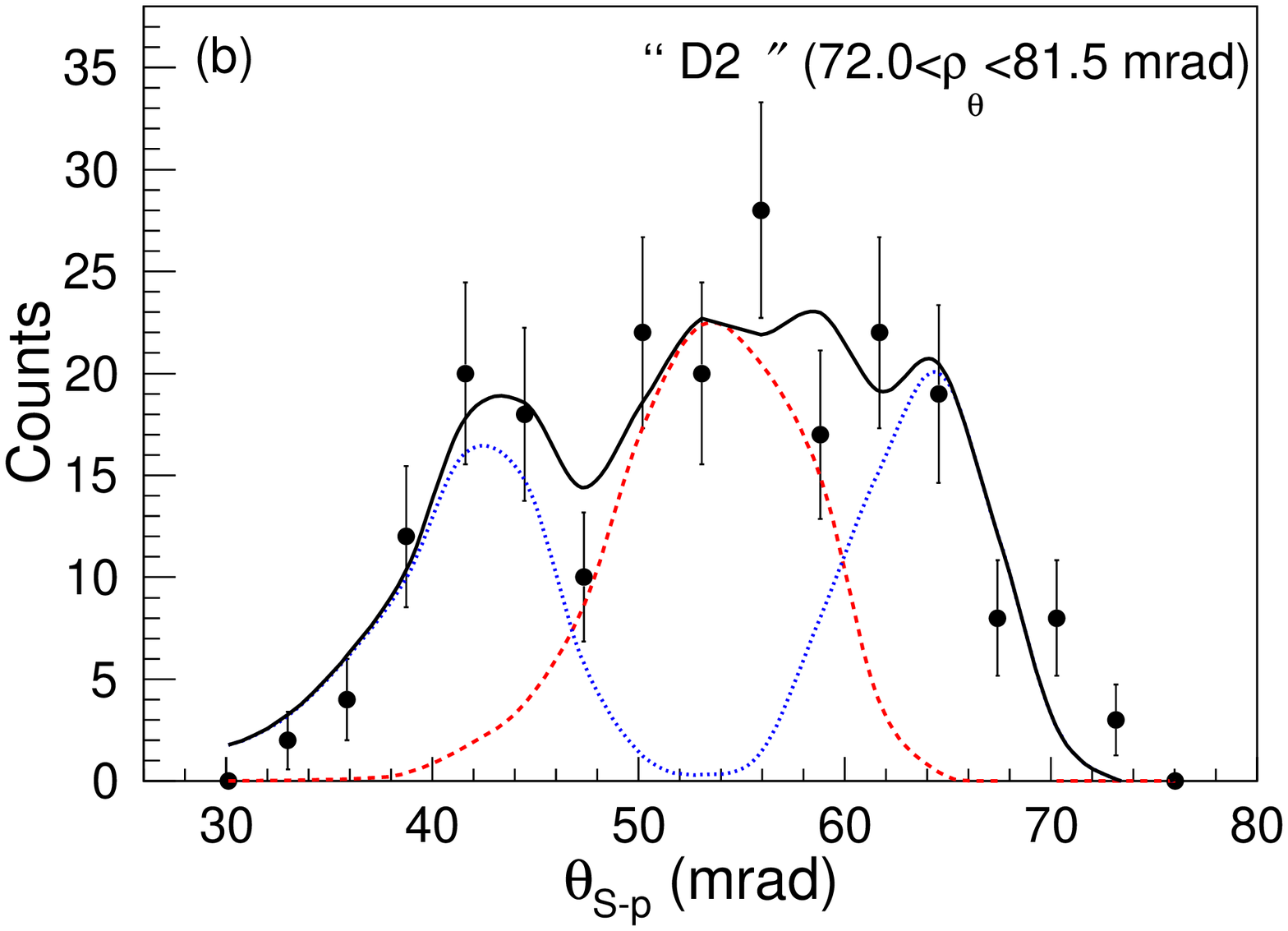}
\end{minipage}
\caption{(Color online) Angular $\theta_{\rm{S}\text{-}p}$ distributions derived from the decays of $^{30}$Ar excited states at 3.9 MeV and at 4.2 MeV above the 2\emph{p} threshold. (a) The data (black dots with statistical uncertainties) are selected from the $^{28}\rm{S}+\emph{p}+\emph{p}$ coincidences by using the $\rho_\theta$ gate ``D1'' at $67.0<\rho_\theta<72.0$ mrad. The solid curve displays the simulation of the sequential 2\emph{p} decay of the $^{30}$Ar state at 3.9 MeV via the $^{29}$Cl resonance at 1.8 MeV (dashed curve) and the 2.9 MeV state (dotted curve). (b) 2\emph{p} decays selected by the $\rho_\theta$ gate ``D2'', $72.0<\rho_\theta<81.5$ mrad. The solid curve is the simulation of the sequential 2\emph{p} decay of the $^{30}$Ar state at 4.2 MeV state via the 2.3 MeV (dashed curve) and the 2.9 MeV (dotted curve) levels in $^{29}$Cl.}
\label{Ar30_D1_D2}
\end{figure}

In order to establish decay mechanisms of the $^{30}$Ar states located above the state ``C'', we inspected the $\theta_{\rm{S}\text{-}p}$ distribution resulting from the decays of such states by imposing the respective arc $\rho_\theta$ gates on the $^{30}$Ar 2\emph{p}-decay events. In Fig.~\ref{Theta_Ar_all}(d1), the $\theta_{\rm{S}\text{-}p}$ spectrum derived from the peak ``D1'' exhibits a triple-peak structure, in which the middle and the right-most peaks match the 1.8 MeV (peak (1) in Fig.~\ref{Theta_Ar_23}) and 2.9 MeV (peak (3) in Fig.~\ref{Theta_Ar_23}) states observed in $^{29}$Cl, respectively. Therefore, a natural interpretation for the experimental $\theta_{\rm{S}\text{-}p}$ spectrum is the sequential proton emission of the $^{30}$Ar state ``D1'' via the above-mentioned two $^{29}$Cl states. To test such an explanation, MC simulations were performed and the resulting $\theta_{\rm{S}\text{-}p}$ spectra were compared with the data displayed in Fig.~\ref{Ar30_D1_D2}(a). There the dashed and dotted curves represent the simulations of the detector response to the 2\emph{p} decay of the 3.9 MeV $^{30}$Ar state via the $^{29}$Cl resonances at 1.8 MeV and 2.9 MeV, respectively. The weighted sum of these two components is shown by the solid curve, and the contributions of the 1.8 MeV and 2.9 MeV components are 60\% and 40\%, respectively. One can see that simulations with our hypothesis reproduce the data rather well. In a similar manner, the $^{28}$S-\emph{p} angular correlations from the peak ``D2'' shown in Fig.~\ref{Theta_rho_Ar} were analyzed. It was found that this $^{30}$Ar state decays by sequential 2\emph{p} emission via the $^{29}$Cl states at 2.3 MeV and 2.9 MeV, respectively. Corresponding MC simulations are shown in Fig.~\ref{Ar30_D1_D2}(b) and reproduce data well.

\begin{figure*}[!htbp]
\centering
\begin{minipage}[b]{0.45\linewidth}
\centering
\includegraphics[scale=0.38, angle=0]{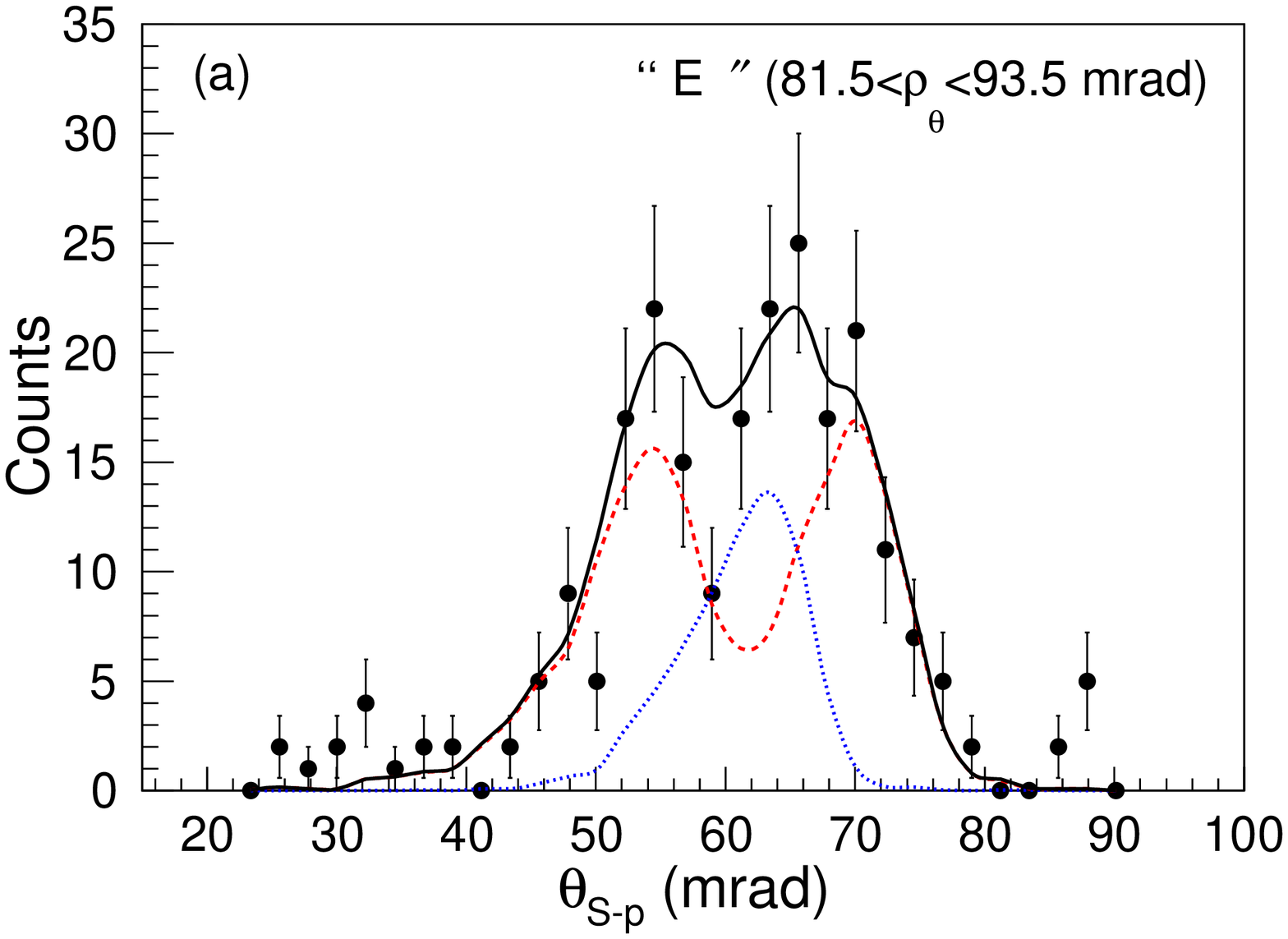}
\end{minipage}
\begin{minipage}[b]{0.45\linewidth}
\centering
\includegraphics[scale=0.38, angle=0]{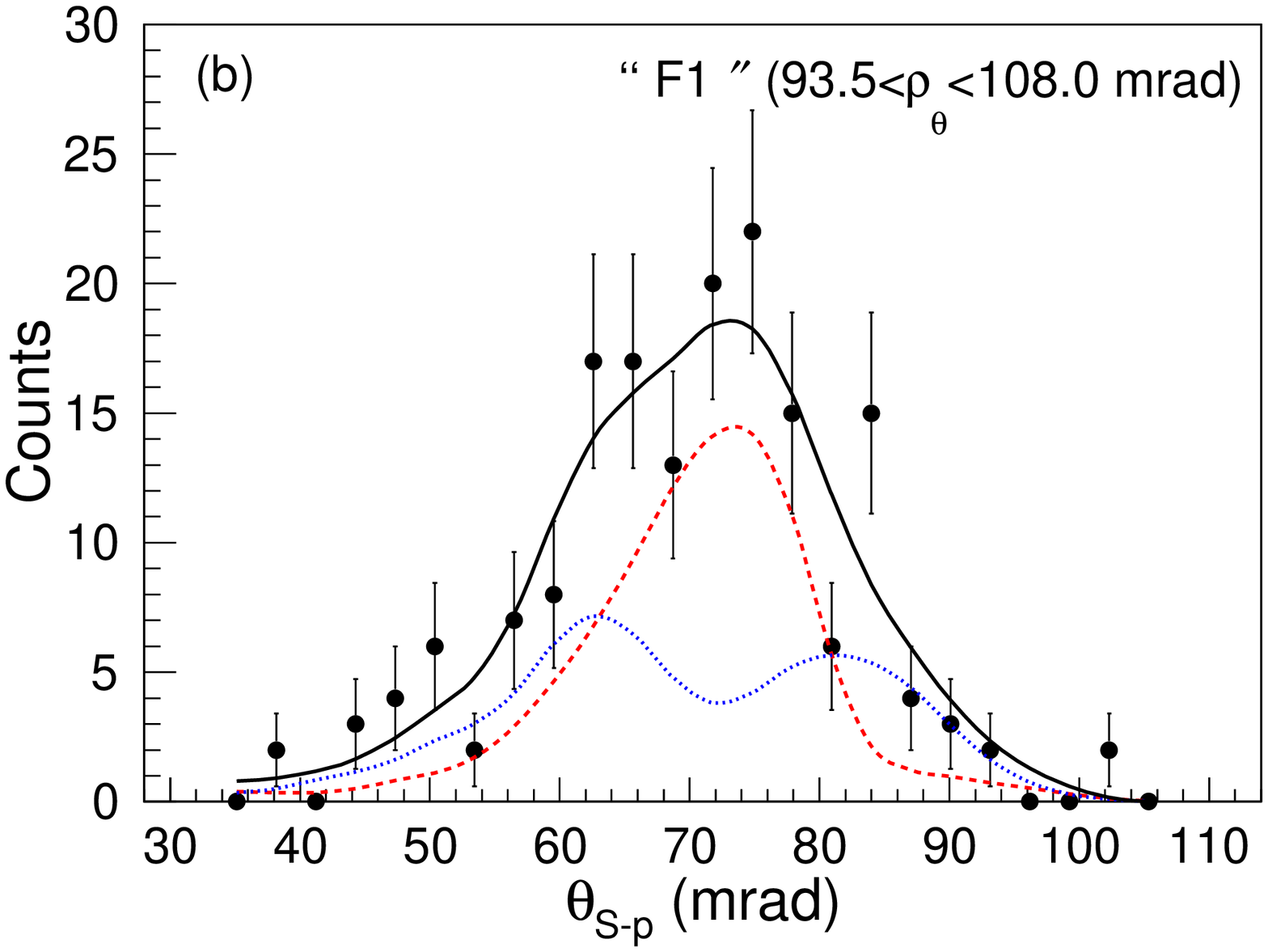}
\end{minipage}
\centering
\begin{minipage}[b]{0.45\linewidth}
\centering
\includegraphics[scale=0.38, angle=0]{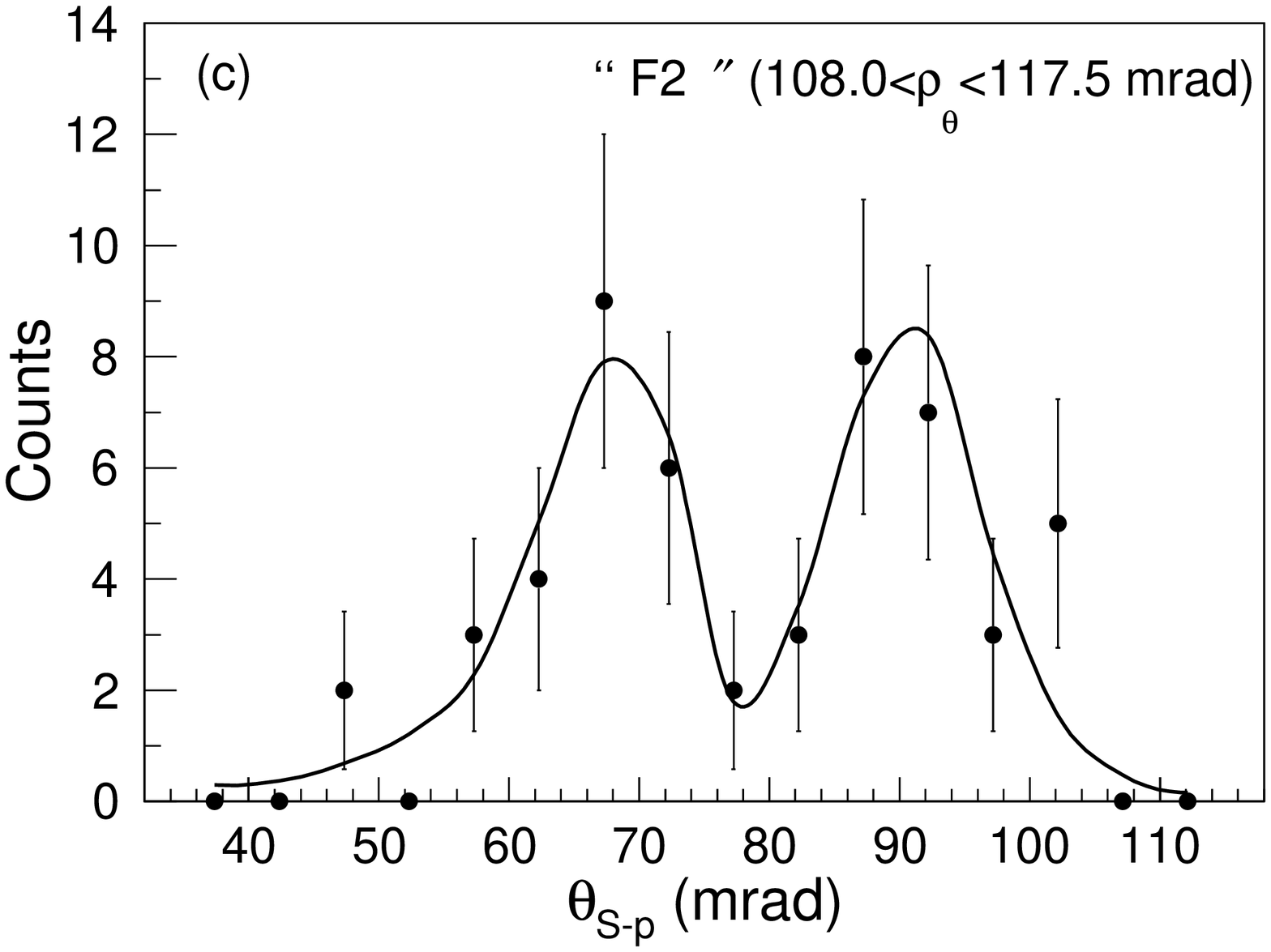}
\end{minipage}
\begin{minipage}[b]{0.45\linewidth}
\centering
\includegraphics[scale=0.38, angle=0]{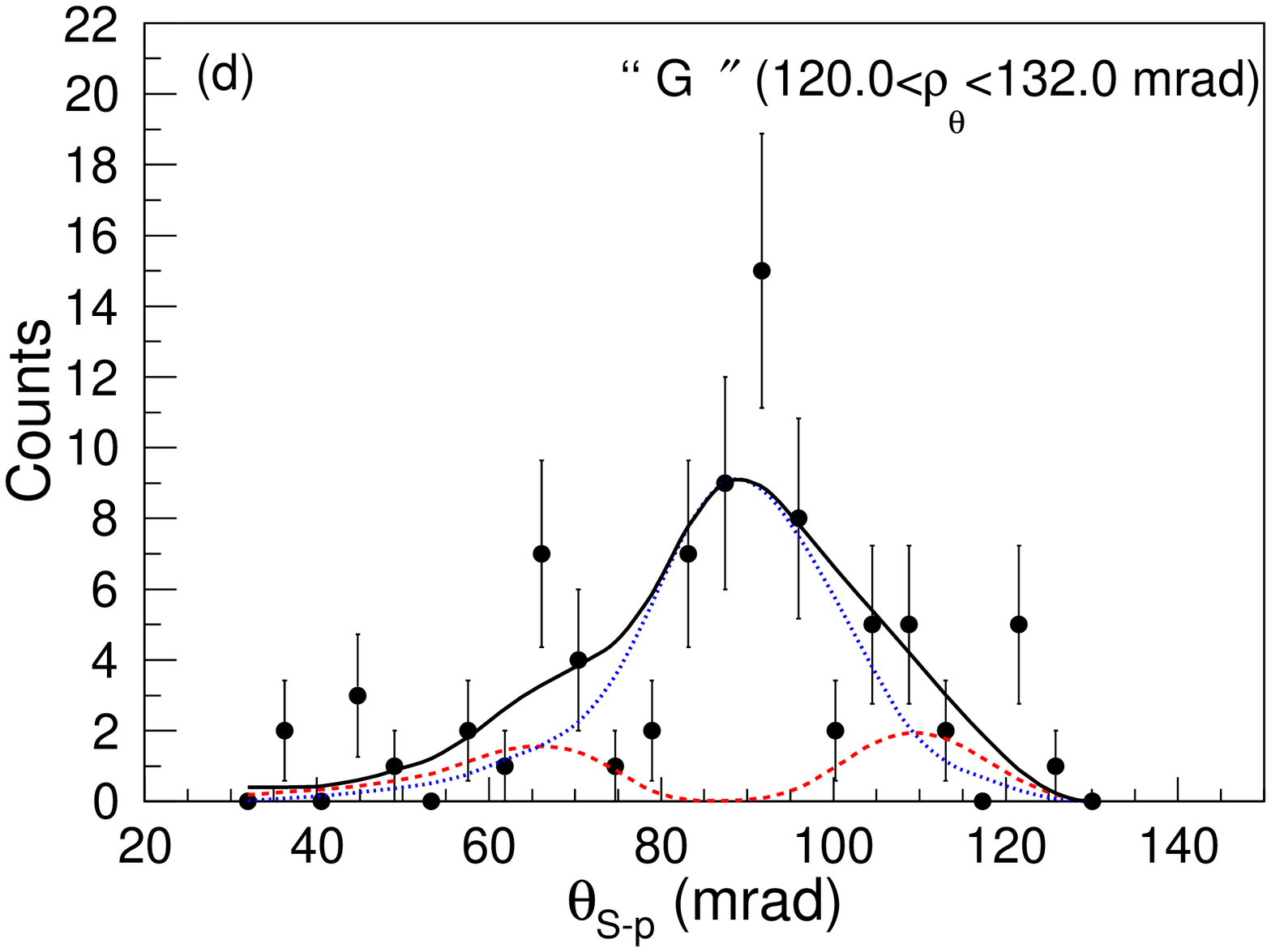}
\end{minipage}
\caption{(Color online) Angular $\theta_{\rm{S}\text{-}p}$ distributions reflecting the decays of several excited states in $^{30}$Ar. (a) 2\emph{p} decays selected by the $\rho_\theta$ gate ``E'', $81.5<\rho_\theta<93.5$ mrad. The simulation of sequential 2\emph{p} emission from the $^{30}$Ar e.s~at 5.6 MeV via the 3.5 MeV (dashed curve) and the 2.9 MeV (dotted curve) states in $^{29}$Cl
is depicted by the solid curve. (b) 2\emph{p} decays of excited state in $^{30}$Ar selected by the $\rho_\theta$ gate ``F1'', $93.5<\rho_\theta<108.0$ mrad. The dashed and dotted curves are the $\theta_{\rm{S}\text{-}p}$ distributions
obtained by simulations of 2\emph{p} emission of the $^{30}$Ar state at 7.9 MeV via two $^{29}$Cl states at 3.9 MeV and 2.9 MeV, respectively. The solid curve represents the summed fit. (c) Data obtained by imposing the $\rho_\theta$ gate ``F2'', $108.0<\rho_\theta<117.5$ mrad. The solid curve displays the $\theta_{\rm{S}\text{-}p}$ spectrum obtained from
the simulation of sequential proton ejection from the 9.4 MeV $^{30}$Ar excited state via the $^{29}$Cl resonance at 3.5 MeV. (d) The 2\emph{p} decays selected by the $\rho_\theta$ gate ``G'', $120.0<\rho_\theta<132.0$ mrad. The solid curve shows the simulation of sequential 2\emph{p} emission of the $^{30}$Ar excited state at 12.6 MeV via the $^{29}$Cl states at 3.5 MeV (dashed curve) and 5.3 MeV (dotted curve), respectively.}
\label{Ar_EFG}
\end{figure*}

Regarding other observed excited states of $^{30}$Ar, namely the peaks ``E'', ``F1'', ``F2'', and ``G'' shown in Fig.~\ref{Theta_rho_Ar}, one can clearly see from Fig.~\ref{Theta_Ar_all} that their $^{28}$S-proton angular correlation spectra display multiple-peak structures. For instance, the $\theta_{\rm{S}\text{-}p}$ distribution obtained from $\rho_\theta$ peak ``E'' has a triple-peak pattern, in which the second and third peaks correspond to the 2.9 MeV and 3.5 MeV state of $^{29}$Cl respectively. Naturally, the decay pattern shown in the Fig.~\ref{Theta_Ar_all}(e) can be attributed to the sequential emission of protons of one $^{30}$Ar state via the above-mentioned two $^{29}$Cl states. Such a hypothesis was tested by performing MC simulations of the detector response to the decays of this $^{30}$Ar$^*$ state. The input parameter of the $^{30}$Ar state is $Q_{2p} = 5.6$ MeV. The resonance energies of two states in $^{29}$Cl are 2.9 MeV and 3.5 MeV respectively. The branching ratios for the two above-mentioned decay branches are 30\% and 70\%, respectively. The simulated $\theta_{\rm{S}\text{-}p}$ distributions are displayed by the dotted and dashed curves in Fig.~\ref{Ar_EFG}(a). The solid curve represents the summed fit which reproduces the data quantitatively. Following a similar analysis of decay patterns from other excited states in $^{30}$Ar, one may tentatively suggest that all these excited states decay by sequential emission of protons via intermediate resonances in $^{29}$Cl. In order to verify such a hypothesis, MC simulations of the detector response to the 2\emph{p} decays from these states were performed. Fig.~\ref{Ar_EFG} shows a comparison between the simulated $\theta_{\rm{S}\text{-}p}$ spectra and the respective experimental distributions. One can see that all simulations agree with the data.

\section{Summary}\label{sec:outlook}

The present work has investigated two proton-unbound nuclei $^{30}$Ar and $^{29}$Cl, which were identified by measuring the trajectories of their in-flight decay products $^{28}\rm{S}+\emph{p}+\emph{p}$ and $^{28}\rm{S}+\emph{p}$, respectively.

For calibration purposes, the decays of the previously-known true 2\emph{p} emitter $^{19}$Mg were remeasured. The 2\emph{p} radioactivity of the $^{19}$Mg ground state and the sequential emission of protons from several known excited states in $^{19}$Mg were confirmed. The deduced 2\emph{p} decay energies are consistent with previous data. Evidence for a new excited state in $^{19}$Mg at $8.9^{+0.8}_{-0.7}$ MeV above the 2\emph{p} threshold was found. We tentatively suggest that this new $^{19}$Mg state decays by sequential emission of protons via two so far unknown $^{18}$Na resonances at $2.5^{+0.7}_{-0.3}$ MeV and $4.0^{+1.5}_{-0.6}$ MeV above the 1\emph{p} threshold, respectively.

By analyzing the $^{28}$S-$p$ and $^{28}$S-$p$-$p$ angular correlations, the $^{30}$Ar g.s.\ was found to be located at $2.45^{+0.05}_{-0.10}$ MeV above the 2\emph{p}-emission threshold and the $^{29}$Cl g.s.\ was found to be $1.8 \pm 0.1$ MeV above the \emph{p}-emission threshold. The level and decay schemes of the observed states in $^{30}$Ar and $^{29}$Cl were reconstructed up to 13 and 6 MeV of excitation respectively.

Several problems relevant to the interpretation of the data were also discussed in this work. These include: Thomas-Ehrman shift in the states of $^{29}$Cl and $^{30}$Ar; transition character of decay dynamics of the $^{30}$Ar g.s.\ and the possibility to improve the determination of $^{29}$Cl and $^{30}$Ar ground states properties; evidence for that the structure of the first excited states of $^{29}$Cl and $^{30}$Ar is dominated by the $^{28}$S core in the $2^+$ state; decay schemes of higher excited states of $^{30}$Ar.

The performed experimental studies are on the edge of modern experimental opportunities. Because of the limited statistics of the data, several issues of the corresponding nuclear structure cannot be elaborated completely, which leaves these aspects for future investigations.

\acknowledgments
This work was supported in part by the Helmholtz International Center for FAIR (HIC for FAIR), the Helmholtz Association (grant IK-RU-002), the Russian Ministry of Education and Science (grant No. NSh-932.2014.2), the Russian Science Foundation (grant No. 17-12-01367), the Polish National Science Center (Contract No. UMO-2011/01/B/ST2/01943), the Polish Ministry of Science and Higher Education (Grant No. 0079/DIA/2014/43, Grant Diamentowy), the Helmholtz-CAS Joint Research Group (grant HCJRG-108), the FPA2009-08848 contract (MICINN, Spain), the Justus-Liebig-Universit\"{a}t Gie{\ss}en (JLU) and GSI under the JLU-GSI strategic Helmholtz partnership agreement. This article is a part of PhD thesis of X.-D. Xu. The authors acknowledge the help of D. Kostyleva in the preparation of the manuscript.


\begin{thebibliography}{31}

\bibitem{Goldansky1960NP}
V. Goldansky, Nucl. Phys. {\bf19}, 482 (1960).

\bibitem{Pfutzner2002EPJA}
M. Pf\"utzner, {\it et al.}, Eur. Phys. J. A {\bf14}, 279 (2002).

\bibitem{Giovinazzo2002PRL}
J. Giovinazzo, {\it et al.}, Phys. Rev. Lett. {\bf89}, 102501 (2002).

\bibitem{Blank2005PRL}
B. Blank, {\it et al.}, Phys. Rev. Lett. {\bf94}, 232501 (2005).

\bibitem{Mukha2007PRL}
I. Mukha, {\it et al.}, Phys. Rev. Lett. {\bf99}, 182501 (2007).

\bibitem{Pomorski2011PRC}
M. Pomorski, {\it et al.}, Phys. Rev. C {\bf83}, 061303 (2011).

\bibitem{Goigoux2016PRL}
T. Goigoux, {\it et al.}, Phys. Rev. Lett. {\bf117}, 162501 (2016).

\bibitem{Grigorenko2003PRC}
L. V. Grigorenko and M. V. Zhukov, Phys. Rev. C {\bf68}, 054005 (2003).

\bibitem{Mukha2010PRC}
I. Mukha, {\it et al.}, Phys. Rev. C {\bf82}, 054315 (2010).

\bibitem{Voss2014PRC}
P. Voss, {\it et al.}, Phys. Rev. C {\bf90}, 014301 (2014).

\bibitem{Olsen2013PRL}
E. Olsen, M. Pf\"utzner, N. Birge, M. Brown, W. Nazarewicz, and A. Perhac, Phys. Rev. Lett. {\bf110}, 222501 (2013).

\bibitem{Mukha2015PRL}
I. Mukha, {\it et al.}, Phys. Rev. Lett. {\bf115}, 202501 (2015).

\bibitem{Golubkova2016PLB}
T. Golubkova, X.-D. Xu, L. Grigorenko, I. Mukha, C. Scheidenberger, and M. Zhukov, Phys. Lett. B {\bf762}, 263 (2016).

\bibitem{Geissel1992NIMB}
H. Geissel, {\it et al.}, Nucl. Instrum. Methods Phys. Res., Sect. B {\bf70}, 286 (1992).

\bibitem{Berz1987NIMA}
M. Berz, H. Hoffmann, and H. Wollnik, Nucl. Instrum. Methods Phys. Res. A {\bf258}, 402 (1987).

\bibitem{GICOSY}
URL http://web-docs.gsi.de/~weick/gicosy/.

\bibitem{Stanoiu2008NIMB}
M. Stanoiu, {\it et al.}, Nucl. Instrum. Methods Phys. Res., Sect. B {\bf266}, 4625 (2008).

\bibitem{Xu2016thesis}
X.-D. Xu, Ph.D. thesis, Justus-Liebig-Universit\"{a}t, Gie{\ss}en (2016).

\bibitem{Mukha2012PRC}
I. Mukha, {\it et al.}, Phys. Rev. C {\bf85}, 044325 (2012).

\bibitem{Fortune2007PRC}
H. T. Fortune and R. Sherr, Phys. Rev. C {\bf76}, 014313 (2007).

\bibitem{Basunia2013NDS}
M. S. Basunia, Nucl. Data Sheets {\bf114}, 1189 (2013).

\bibitem{Brown2015PRC}
K. W. Brown, {\it et al.}, Phys. Rev. C {\bf92}, 034329 (2015).

\bibitem{Kondo2016PRL}
Y. Kondo, {\it et al.}, Phys. Rev. Lett. {\bf116}, 102503 (2016).

\bibitem{Comay1988PLB}
E. Comay, I. Kelson, and A. Zidon, Phys. Lett. B {\bf210}, 31 (1988).

\bibitem{NNDC}
URL http://www.nndc.bnl.gov.

\bibitem{Brown2006PRC}
B. A. Brown and W. A. Richter, Phys. Rev. C {\bf74}, 034315 (2006).

\bibitem{Thomas1952PR}
R. G. Thomas, Phys. Rev. {\bf88}, 1109 (1952).

\bibitem{Ehrman1951PR}
J. B. Ehrman, Phys. Rev. {\bf81}, 412 (1951).

\bibitem{Pfutzner2012RMP}
M. Pf\"utzner, M. Karny, L. V. Grigorenko, and K. Riisager, Rev. Mod. Phys. {\bf84}, 567 (2012).

\bibitem{Schwierz2007arXiv}
N. Schwierz, I. Wiedenhover, and I. A. Volya, arXiv: 0709.3525 (2007).

\bibitem{Angeli2013ADNDT}
I. Angeli and K. Marinova, At. Data Nucl. Data Tables {\bf99}, 69 (2013).

\end{thebibliography}

\end{document}